\documentclass[a4paper,fleqn,longmktitle]{cas-sc}
\usepackage[justification=centering]{caption}
\usepackage{graphicx} 
\usepackage{float}
\usepackage{cas-common}
\usepackage[numbers]{natbib}

\usepackage{longtable}
\usepackage{booktabs}

\begin{document}
\captionsetup[figure]{labelfont={bf},name={Fig.},labelsep=period}
\let\WriteBookmarks\relax
\def\floatpagepagefraction{1}
\def\textpagefraction{.001}
\shorttitle{A Survey of Recommender Systems Based on Generative Adversarial Networks}
\shortauthors{Min Gao et~al.}

\title [mode = title]{Recommender Systems Based on Generative Adversarial Networks: A Problem-Driven Perspective}                      



\author[1, 2]{Min Gao}

\cormark[1]
\fnmark[1]
\ead{gaomin@cqu.edu.cn}
\address[1]{Key Laboratory of Dependable Service Computing in Cyber Physical Society (Chongqing University), Ministry of Education, Chongqing, 401331, China}
\address[2]{School of Big Data and Software Engineering, Chongqing University, Chongqing, 401331, China}

\author[1, 2]{ Junwei Zhang}
\fnmark[1]

\author[3]{ Junliang Yu}
\address[3]{School of Information Technology and Electrical Engineering, The University of Queensland, Queensland, 4072, Australia}

\author[4]{ Jundong Li}
\address[4]{Department of Electrical and Computer Engineering, Department of Computer Science, and School of Data Science, University of Virginia, Charlottesville, 22904, USA}

\author[1, 2]{ Junhao Wen}

\author[1, 2]{ Qingyu Xiong}

\cortext[cor1]{Corresponding author}
\fntext[myfootnote]{The first two authors contributed equally and are joint first authors.}


\begin{abstract}
Recommender systems (RSs) now play a very important role in the online lives of people as they serve as personalized filters for users to find relevant items from an array of options. Owing to their effectiveness, RSs have been widely employed in consumer-oriented e-commerce platforms. However, despite their empirical successes, these systems still suffer from two limitations: data noise and data sparsity. In recent years, generative adversarial networks (GANs) have garnered increased interest in many fields, owing to their strong capacity to learn complex real data distributions; their abilities to enhance RSs by tackling the challenges these systems exhibit have also been demonstrated in numerous studies. In general, two lines of research have been conducted, and their common ideas can be summarized as follows: (1) for the data noise issue, adversarial perturbations and adversarial sampling-based training often serve as a solution; (2) for the data sparsity issue, data augmentation--implemented by capturing the distribution of real data under the minimax framework--is the primary coping strategy. To gain a comprehensive understanding of these research efforts, we review the corresponding studies and models, organizing them from a problem-driven perspective. More specifically, we propose a taxonomy of these models, along with their detailed descriptions and advantages. Finally, we elaborate on several open issues and current trends in GAN-based RSs.
\end{abstract}



\begin{keywords}
Recommender Systems \sep Generative Adversarial Networks \sep Data Sparsity \sep Data Noise
\end{keywords}

\maketitle

\section{Introduction}
Owing to the rapid development of Internet-based technologies, the quantities of data on the Internet are growing exponentially; as a result, Internet users are persistently inundated with excessive amounts of information \cite{DBLP:journals/isci/WangGWZ20, DBLP:journals/isci/DauSRO20}. In particular, when it comes to online shopping, people struggle to make choices when a vast range of options are presented. As an effective tool to tackle such information overloads, recommender systems (RSs) have been widely used in various online scenarios, including E-commerce (e.g., Amazon and Taobao), music playback (e.g., Pandora and Spotify), movie recommendation (e.g., Netflix and iQiyi), and news recommendation (e.g., BBC News and Headlines).

Despite their pervasiveness and strong performances, RSs still suffer from two main problems: data noise and data sparsity. As an extrinsic problem, data noise stems from the casual, malicious, and uninformative feedback in the training data \cite{DBLP:conf/ijcai/FanD0WTL19, DBLP:conf/sigir/0001HDC18}. More specifically, users occasionally select products outside their interests, and this casual feedback is also collected and indiscriminately used during the RS model training. During training, it is common for the randomly selected negative samples to become uninformative and mislead the recommendation models. After training, a small number of malicious profiles or feedbacks are intermittently injected into the RS, to manipulate recommendation results. Failure to identify these data may result in security problems. Compared with the data noise problem, data sparsity is an intrinsic problem, it occurs because each user generally only consumes a small proportion of the available items. Existing RSs typically rely on the historical interactive information between users and items when capturing the interests of users. When a vast quantity of the data are missing, the RS commonly fails to satisfy users, by making inaccurate recommendations \cite{DBLP:conf/icdm/HeCZC18}. Without coping mechanisms, these problems often cause the RS to fail, leading to an inferior user experience.

Numerous researchers are aware of the negative effects of these two problems and have attempted to minimize the adverse factors. For the data noise problem, a number of solutions have been proposed \cite{DBLP:journals/kbs/ZhangZZW18, DBLP:journals/dss/BagKAT19, DBLP:conf/cikm/Yu0LYL18, DBLP:conf/kdd/ChenSSH17}. Among them, Zhang et al. \cite{DBLP:journals/kbs/ZhangZZW18} applied a hidden Markov model to analyze preference sequences; subsequently, they utilized hierarchical clustering to distinguish attack users from rating behaviors. Bag et al. \cite{DBLP:journals/dss/BagKAT19} proposed a method that corrects casual noise using the Bhattacharya coefficient and concept of self-contradiction. To distinguish more informative items from unobserved ones, Yu et al. \cite{DBLP:conf/cikm/Yu0LYL18} identified more informative users by modeling all training data as a heterogeneous information network, to obtain the embedding representation. To obtain informative negative items, several researchers adopted popularity biased sampling strategies \cite{DBLP:conf/kdd/ChenSSH17}. Although using these methods, it was found that conventional RS models are susceptible to noise in the training data, they can only prevent conspicuous noises from a specific perspective, and they cannot continuously update their ability to manage unobserved patterns of noise. For the data sparsity problem, numerous methods have been developed to integrate a wealth of auxiliary information into the RS. Such data comprise the side information of users and items, as well as the relationships between them \cite{DBLP:conf/sigir/Wang0NC17, DBLP:journals/ijon/XiongQHXBLYY20, DBLP:conf/www/ChengDZK18, DBLP:conf/aaai/WangWX00C19}. For example, Cheng et al. \cite{DBLP:conf/www/ChengDZK18} used textual review information to model user preferences and item features and tackle the problems of data sparsity. Wang et al. \cite{DBLP:conf/aaai/WangWX00C19} extracted auxiliary information from a knowledge graph to enhance recommendations. Although the integration of auxiliary information is useful, these methods still struggle to obtain a satisfactory result, owing to inconsistent data distributions or patterns and large computational costs.

Recently, generative adversarial networks (GANs) have led to rapid developments in deep learning fields \cite{DBLP:conf/cvpr/ChoiCKH0C18, DBLP:journals/tomccap/NieWLNS20, DBLP:journals/taslp/LiuFQL19, DBLP:conf/aims2/WangOTL18}, including image and audio generation \cite{DBLP:journals/tomccap/NieWLNS20, DBLP:journals/taslp/LiuFQL19, DBLP:conf/aims2/WangOTL18}. The principle of GANs is to play an adversarial minimax game between a generator and a discriminator. The generator focuses on capturing the distribution of real observed data, to generate adversarial samples and fool the discriminator; meanwhile, the discriminator attempts to distinguish whether the inputted sample is from the generator or not. This adversarial process continues until the two components reach the Nash equilibrium.

Significant successes realized by applying GANs to deep learning fields have set good examples for RSs, and GAN-based RSs have been introduced in several existing studies \cite{DBLP:conf/ijcai/WuLMZZG19, DBLP:conf/wsdm/BeigiMGAN020, DBLP:conf/ijcai/YuZCX19}. In this study, we identified relevant papers from DBLP using the following keywords: generative adversarial network, GAN, and adversarial. According to the statistics retrieved from top-level conferences related to RSs in the past three years, the number of GAN-based recommendation models is increasing yearly, as shown in Table 1. Meanwhile, in a seminar on GAN-based information retrieval (IR) models presented at SIGIR in 2018, researchers \cite{DBLP:conf/sigir/Zhang18} suggested that GAN-based RSs will become immensely popular in the field of RS research. This is because the GAN concept provides new opportunities to mitigate data noise and data sparsity. Several existing studies have verified the effectiveness of introducing adversarial perturbations and the minimax game into the objective function, to reduce data noise. Other studies have attempted to use the discriminator to recognize informative examples in an adversarial manner. Meanwhile, to address the data sparsity issue, a separate line of research has investigated  the capabilities of GANs to generate user profiles by augmenting user-item interaction and auxiliary information.

\begin{table}[pos=htp]

 \centering{ 
      \begin{bfseries}   
         Table 1
      \end{bfseries}
      Statistics of papers describing GAN-based recommendation models}

\begin{tabular}{l|lll}
\hline
\textbf{Meeting\textbackslash{}Year} & \textbf{2017} & \textbf{2018} & \textbf{2019} \\ \hline
\textbf{AAAI}                        & 0             & 1             & 1             \\
\rowcolor[HTML]{EFEFEF} 
\textbf{CIKM}                        & 0             & 2             & 2             \\
\textbf{ICML}                        & 0             & 0             & 1             \\
\rowcolor[HTML]{EFEFEF} 
\textbf{IJCAI}                       & 0             & 1             & 3             \\
\textbf{KDD}                         & 0             & 1             & 3             \\
\rowcolor[HTML]{EFEFEF} 
\textbf{RecSys}                      & 0             & 2             & 3             \\
\textbf{SIGIR}                       & 1             & 4             & 4             \\ \hline
\rowcolor[HTML]{EFEFEF} 
\textbf{Total}                       & 1             & 11            & 17            \\ \hline
\end{tabular}
\end{table}

To the best of our knowledge, only a small number of systematic reviews have sufficiently analyzed existing studies and the current progress of GAN-based recommendation models. To this end, we investigate and review various GAN-based RSs from a problem-driven perspective. More specifically , we classify the existing studies into two categories: the first reviews models designed to reduce the adverse effects of data noise; the second focuses on the models designed to mitigate the data sparsity problem. We hope that this review will lay the foundation for subsequent research on GAN-based RSs. The primary contributions of this review are summarized as follows:

\begin{enumerate}[\textbullet]
\item To gain a comprehensive understanding of the state-of-the-art GAN-based recommendation models, we provide a retrospective survey of these studies and organize them from a problem-driven perspective.
\item We systematically analyze and investigate the capabilities of GAN-based models to mitigate data noise issues arising from two different sources: (1) models to mitigate casual and malicious noise, and (2) models to distinguish informative samples from unobserved items.
\item We conduct a systematic review of recommendation models that implement GANs to alleviate data sparsity issues arising from two different sources: (1) models for generating user preferences through augmentation with interactive information, and (2) models for synthesizing user preferences through augmentation with auxiliary information.
\item We elaborate on several open issues and current trends in GAN-based RSs.
\end{enumerate}

The remainder of this paper is organized as follows. The development process of GANs is described in Section 2. In Sections 3 and 4, up-to-date GAN-based recommendation models are introduced in a problem-driven way, highlighting the efforts devoted to mitigating the problems of data noise and data sparsity, respectively. In Section 5, we discuss prominent challenges and research directions. In the final section, we present the conclusions of our work.

\section{The Foundations of Generative Adversarial Networks}
In this section, we first summarize some common notations in GANs, to simplify the subsequent GAN-related content. Subsequently, we present some classical GANs.

\subsection{Notations}
Throughout this paper, $G$ denotes the generator of a GAN, whilst $D$ denotes its discriminator. $\mathbb{E}$ indicates the expectation calculation. $P$ represents the data distribution. $L$ denotes the loss function of the introduced model. These notations are summarized in Table 2.

\begin{table*}[pos=htp]

 \centering{ 
      \begin{bfseries}   
         Table 2
      \end{bfseries}
      Notations used throughout this paper.\\}
\begin{tabular}{l|l}
\hline
\multicolumn{1}{c|}{\textbf{Notation}} & \multicolumn{1}{c}{\textbf{Description}} \\ \hline
$G$                                      & The generator                            \\
\rowcolor[HTML]{EFEFEF} 
$D$                                      & The discriminator                          \\
$\mathbb{E}$                             & The exception value                      \\
\rowcolor[HTML]{EFEFEF} 
$L$                                      & The loss function                       \\
$P$                                      & The distribution                         \\ \hline
\end{tabular}
\end{table*}

\subsection{Typical Model}
The GAN, which is an unsupervised model proposed by Goodfellow et al. in 2014 \cite{DBLP:conf/nips/GoodfellowPMXWOCB14}, has attracted widespread attention from both academia and industry. It features two components: a generator and a discriminator. The former learns to generate data that conform to the distribution of real data as much as possible; the latter must distinguish between the real data and those generated by the generator. The two models compete against each other and optimize themselves via feedback loops. The process is shown in Fig. 1.

\begin{figure}[pos=htp]
  \centering
    \includegraphics{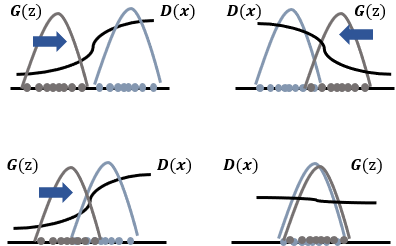}
    
  \centering{ 
      \begin{bfseries}   
          Fig. 1.
      \end{bfseries}
      Basic GAN concept.}
\end{figure}

In Fig. 1, $z$ denotes the random noise, and $x$ denotes the real data. The loss function is

\begin{equation}
\label{eq:eq1}
\begin{aligned} \min _{G} \max _{D} L(D, G) = \mathbb{E}_{x \sim p_{data}(x)}[\log D(x)] + \mathbb{E}_{z \sim p_{z}(x)}[\log (1-D(G(z)))], \end{aligned}
\end{equation}where $p_{data}$ is the probability distribution of the generated data. These models---the so-called ``vanilla GANs''---exhibit several shortcomings. They cannot indicate the training process of their loss functions and lack diversity in their generated samples. Martin et al. \cite{DBLP:conf/icml/ArjovskyCB17} have investigated the causes of these flaws; they found that when the distributions of real and generated data are non-overlapping, the function tends to be constant, which leads to the disappearance of the gradient. Then, they \cite{DBLP:conf/iclr/ArjovskyB17} proposed Wasserstein GAN (WGAN), which uses the Wasserstein distance instead of the original Jensen-Shannon divergence. $f_{w}$ is the function that calculates the Wasserstein distance. The loss function of a WGAN is

\begin{equation}
\label{eq:eq2}
L^{D}= \mathbb{E}_{z \sim P_{z}}\left[f_{w}(G(z))\right]-\mathbb{E}_{x \sim P_{r}}\left[f_{w}(x)\right].
\end{equation} 

Although WGANs can theoretically mitigate the problems caused by training difficulties, they still suffer from their own problems: the generated samples are of low quality, and the training fails to converge owing to the Lipschitz constraint on the discriminator. Consequently, Ishaan et al. \cite{DBLP:conf/nips/GulrajaniAADC17} regularized the Lipschitz constraints and proposed a gradient penalty WGAN model (WGAN-GP). The Lipschitz constraint was approximated by assigning the constraint to the penalty term of the objective function. The loss function is

\begin{equation}
\label{eq:eq3}
\begin{aligned} L= \mathbb{E}_{\tilde{x} \sim P_{g}}[D(\tilde{x})]-\mathbb{E}_{x \sim P_{r}}[D(x)] + \omega \mathbb{E}_{\hat{x} \sim P_{\tilde{x}}}\left[\left(\left\|\nabla_{\hat{x}} D(\hat{x})\right\|_{2}-1\right)^{2}\right], \end{aligned}
\end{equation}where $P_{r}$ is the data distribution and $P_{g}$ the generator distribution. As shown in Eq. \ref{eq:eq3}, the larger the parameter $x$, the smoother the log loss function, and the smaller the gradient; this leads to almost no improvement in the generator. Mao et al. \cite{DBLP:conf/iccv/MaoLXLWS17} proposed the least-squares GAN (LSGAN); this was inspired by WGAN and provides the discriminator $D$ with a loss function for smooth and unsaturated gradients. LSGAN alleviates the instability of training and improves the quality of the generated data. 

In addition to modifying the loss function to improve the performance of GANs, several studies \cite{DBLP:journals/corr/RadfordMC15, DBLP:conf/nips/DentonCSF15, DBLP:conf/nips/ChenCDHSSA16} have focused on the network structure of the discriminator and generator. The structures of vanilla GANs are realized using a multi-layer perceptron (MLP), which makes parameter tuning difficult. To solve this problem, Alec et al. \cite{DBLP:journals/corr/RadfordMC15} proposed a deep convolutional GAN (DCGAN), because convolutional neural networks (CNNs) exhibit superior capacities for fitting and representing compared to MLPs. DCGAN dramatically enhanced the quality of data generation and provided a reference on neural network structures for subsequent research into GANs. The framework of the DCGAN is shown in Fig. 2.

\begin{figure}[pos=htp]
  \centering
    \includegraphics{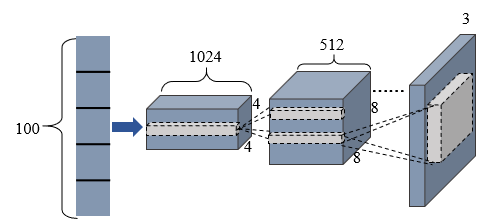}
    
  \centering{ 
      \begin{bfseries}   
          Fig. 2.
      \end{bfseries}
      DCGAN framework \cite{DBLP:journals/corr/RadfordMC15}.}
\end{figure}

Although GANs have received considerable attention as unsupervised models, the GAN generator can only generate data based on random noise, which renders the generated data unusable. Therefore, Mirza and Osindero \cite{DBLP:journals/corr/MirzaO14} proposed the conditional GAN (CGAN). By adding conditional constraints to the model, the CGAN generator was able to generate condition-related data. CGAN can be seen as an enhanced model that converts an unsupervised GAN into a supervised one. This improvement has been proven effective and guided subsequent related work. For instance, the Laplacian pyramid of adversarial networks (LAPGAN) \cite{DBLP:conf/nips/DentonCSF15} integrates a CGAN within the framework of the Laplacian pyramid, to generate coarse-to-fine fashion images. InfoGAN \cite{DBLP:conf/nips/ChenCDHSSA16} is another variant of CGANs; it decomposes random noise into noise  and implicit encoding to learn more interpretable and meaningful representations.

After considering the current progress of GANs in detail, we find that many advanced models \cite{DBLP:conf/acml/WuXTZQLL18, DBLP:conf/cvpr/ChoiCKH0C18, DBLP:journals/tomccap/NieWLNS20, DBLP:journals/taslp/LiuFQL19} have been specifically proposed in the fields of computer vision and natural language processing. These models do have some reference value for mitigating the data noise and sparsity problems of RSs. We introduce GAN-based recommendation models in the next two sections.

\section{GAN-Based Recommendation Models for Mitigating the Data Noise Issue}
In RS research, the problem of data noise is attracting increasing amounts of attention. It affects not only the accuracy of RSs but also their robustness. In this section, we review the state-of-the-art GAN-based models designed to identify casual and malicious noise and uninformative feedback. We categorize the GAN-based models into two categories, in terms of the sources of data noise described in the introduction: (1) models for mitigating casual and malicious noise, and (2) models for distinguishing informative samples from unobserved items.

\subsection{Models for Mitigating Casual and Malicious Noise}
[\textbf{APR}] Applying adversarial learning to the construction processes of recommendation models is a common method of mitigating the problem of data noise, which includes casual and malicious noise. He et al. \cite{DBLP:conf/sigir/0001HDC18} were the first to verify the effectiveness of adding adversarial perturbations into RSs, and they proposed an adversarial personalized ranking (APR) model to enhance model generalizability. Furthermore, the purpose of adversarial perturbations is to help the model preemptively consider the bias caused by noise. Specifically, the loss function of the APR model contains two parts: one adds perturbations to the parameters of the Bayesian personalized ranking model (BPR) and makes the performance as low as possible; the other, without adversarial perturbations, makes the recommendation performance as high as possible. The loss function of the APR model is composed of these two parts, as shown in Eq. 4:

\begin{equation}
\label{eq:eq4}
\begin{aligned}
L_{APR}\left ( D| \Theta \right ) =  L_{BPR}\left ( D| \Theta \right ) + \omega L_{BPR}\left ( D| \Theta+\Delta_{adv} \right ),
\end{aligned}
\end{equation}

\begin{equation}
\label{eq:eq5}
\Delta_{adv}= arg \max _{\Delta,\left \| \Delta \leq \epsilon \right \|} L_{BPR}(D| \hat{\Theta}+\Delta_{adv}),
\end{equation}where $\Delta$ represents the disturbance of the model parameters, $0 \leq \epsilon$ controls the magnitude of the disturbance, $\hat{\Theta}$ represents the parameters of the existing model, and $\omega$ is the equilibrium coefficient that controls the strength of the regularization term $L_{BPR}\left ( D| \Theta+\Delta_{adv} \right )$. In contrast to BPR, APR performs adversarial training using the fast gradient method \cite{DBLP:journals/corr/GoodfellowSS14}, finding the optimal perturbations and parameters to alleviate the data noise problem.

This is an innovative idea, because it uses adversarial perturbations to simulate malicious noise and thereby improve the robustness of the BPR model; this demonstrates the competitive results that can be obtained by applying adversarial training to BPR. In ranking tasks, APR models have achieved better recommendation performances than deep neural network (DNN) based RSs \cite{DBLP:conf/www/HeLZNHC17}, and they have subsequently inspired many research works. 

\begin{figure}[pos=htp]
  \centering
    \includegraphics{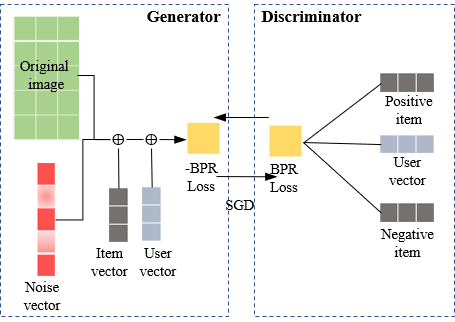}
    
  \centering{ 
      \begin{bfseries}   
          Fig. 3.
      \end{bfseries}
      AMR framework \cite{DBLP:journals/corr/abs-1809-07062}.}
\end{figure}

[\textbf{AMR}] By extending APR, Tang et al. \cite{DBLP:journals/corr/abs-1809-07062} devised an adversarial multimedia recommendation (AMR) model for image RSs. Based on the visualized BPR (VBPR) \cite{DBLP:conf/aaai/HeM16}, AMR adds adversarial perturbations to the low-dimensional features of images extracted by a CNN, to learn robust image embeddings. The loss function of AMR is

\begin{equation}
\label{eq:eq6}
\begin{aligned}
\theta^{*},\Delta ^{*}=  arg \min _{\theta } \max _{\Delta} (L_{BPR}\left ( \theta  \right ) + \omega L^{'}_{BPR}\left ( \theta ,\Delta  \right )),
\end{aligned}
\end{equation}

\begin{equation}
\label{eq:eq7}
\widehat{y}^{'}_{ui}=p_{u}^{T}\left ( q_{i}+ E\cdot \left ( c_{i}+\Delta _{i} \right ) \right ),
\end{equation}

\begin{equation}
\label{eq:eq8}
\Delta ^{*}=arg max_{\Delta }=\sum_{(u,i)\in D}-ln\sigma \left ( \hat{y}^{'}_{ui} - \hat{y}^{'}_{uj} \right ),
\end{equation}where $\Delta_{i}$ represents the adversarial perturbations and the optimal perturbation is obtained by maximizing the loss function in the training data. $\theta$ represents the parameters in AMR. Besides this, Tran et al. \cite{DBLP:conf/sigir/TranSL19} used APR as a component of a music sequence recommendation model, to improve its robustness. However, it is difficult to derive exact optimal perturbations in the above models. 

[\textbf{MASR}] Besides AMR, Tran et al. \cite{DBLP:conf/sigir/TranSL19} used APR as a component of a music sequence recommendation model, to improve its robustness. However, it is difficult to derive exact optimal perturbations in the above models.

[\textbf{CollaGAN}] Tong et al. \cite{DBLP:conf/icde/TongLZSC19} proposed a collaborative GAN (CGAN ) to mitigate the adverse impacts of noise and improve the robustness of RSs. More specifically, CGAN uses a variational auto-encoder \cite{DBLP:journals/corr/KingmaW13} as a generator, and its input is the rating vector for each item. After encoding, the generator learns the data distribution from the training data and generates fake samples through the embedding layer. The discriminator focuses on maximizing the likelihood of distinguishing generated item samples from real item vectors. The loss functions of the generator and discriminator are

\begin{equation}
\label{eq:eq9}
L^{G}=-\mathbb{E}_{i \sim P_{\theta}(i | u)}[D(v | u)],
\end{equation}

\begin{equation}
\label{eq:eq10}
\begin{aligned}
L^{D}= -\mathbb{E}_{v \sim P_{r}(v|u)}[D(v | u)] + \mathbb{E}_{x \sim P_{\theta}(v | u)}[D(v | u)]  + \omega \mathbb{E}_{j \sim P_{\vec{p}}(\hat{v}|u)}\left[\left(\left\|\nabla_{\hat{v}} D(\hat{v}|u)\right\|_{2}-1\right)^{2}\right],
\end{aligned}
\end{equation}respectively. Here, $u$ and $v$ represent the low-dimensional vectors of the user and item, respectively. In addition, the authors also developed the vanilla GAN into a WGAN and WGAN-GP, to exploit the faster training speeds and superior performances of these models. Compared with other models such as neural matrix factorization (NeuMF) \cite{DBLP:conf/www/HeLZNHC17}, IRGAN  \cite{DBLP:conf/sigir/WangYZGXWZZ17}, and GraphGAN \cite{DBLP:conf/aaai/WangWWZZZXG18}, the performance of CGAN was significantly enhanced for two movie-recommendation datasets \cite{DBLP:journals/tiis/HarperK16, DBLP:journals/computer/KorenBV09}. Moreover, CGAN adopts a conventional point-wise loss function instead of a pair-wise one. The performance of the entire model can be further improved if the optimization component is designed to be more elegant.

[\textbf{ACAE \& FG-ACAE }] Yuan et al. \cite{DBLP:conf/ijcnn/YuanYB19} proposed a general adversarial training framework---referred to as an atrous convolution autoencoder  (ACAE)---for DNN-based RSs. They implemented it with  a collaborative auto-encoder, to strike a balance between accuracy and robustness by adding perturbations at different parameter levels. The framework of the ACAE is shown in Fig. 4. The experimental results confirmed that adding perturbations has a more significant impact on the original model, where the effect of adding perturbations to the decoder weights---rather than those of the encoder---is higher. The effects of adding perturbations to user-embedding vectors and hidden layers are negligible. To control the perturbations more precisely, they \cite{DBLP:conf/sigir/YuanYB19} used different coefficients to separately control noise terms and obtain further benefits from the adversarial training.

\begin{figure}[pos=htp]
  \centering
    \includegraphics{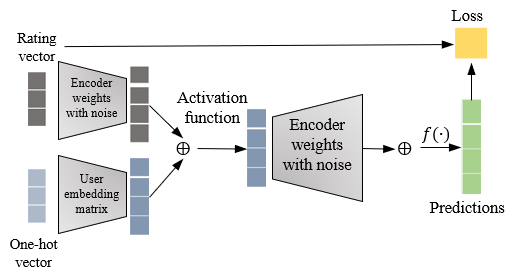}
    
  \centering{ 
      \begin{bfseries}   
          Fig. 4.
      \end{bfseries}
      ACAE framework \cite{DBLP:conf/ijcnn/YuanYB19}.}
\end{figure}

[\textbf{SACRA}] Because of the vulnerabilities of neural networks, the adoption of adversarial learning frameworks has gradually increased in related research. Li et al. \cite{DBLP:conf/wsdm/LiW020} adopted the idea of adversarial perturbations at the embedding level; they proposed a self-attentive prospective customer recommendation model, referred to as SACRA. The model implements the fast gradient method to estimate adversarial perturbations.

[\textbf{ATF}] In addition, adversarial tensor factorization (ATF) was proposed by Chen et al. \cite{DBLP:conf/recsys/ChenL19}; it is considered to be the first attempt to integrate adversarial perturbations into models based on tensor factorization. ATF outperformed state-of-the-art tensor-based RSs. This work can incorporate more contextual data, including location, time, and social information.

\subsection{Models for Distinguishing Informative Samples from Unobserved Items}
It is difficult for RSs to extract informative data from large quantities of unobserved data. More specifically, when models optimize their pair-wise objective functions, the negative sampling technique often provides more informative samples. Hence, it is crucial to provide informative negative samples dynamically. 

[\textbf{IRGAN}] The first GAN designed to mitigate this problem was IRGAN, proposed by Wang et al. \cite{DBLP:conf/sigir/WangYZGXWZZ17}. IRGAN unifies the generative retrieval and discriminative models; the former predicts relevant documents for a given query and the latter predicts the relevancy in each query document pair. The generative model is used as the generator; it selects informative items for a given user by fitting the real relevance distribution over items. The discriminative retrieval model, employed as the discriminator, distinguishes between relevant and selected items. Then, the discriminator feeds the result into the generator, to select more informative items. The generated items are inputted into the discriminator to mislead it. The loss function of IRGAN is

\begin{equation}
\label{eq:eq11}
\begin{aligned} L^{G, D} =  \min _{\theta} \max _{\phi} \sum_{n=1}^{N} \left(\mathbb{E}_{i \sim P_{true }\left(i | u_{n}, r\right)}\left[\log D\left(i | u_{n}\right)\right] + \mathbb{E}_{i \sim P_{\theta}\left(i|u_{n}, r\right)}\left[\log \left(1-D\left(i | u_{n}\right)\right)\right]\right),
\end{aligned}
\end{equation}where $D(i|u)=\frac{\exp \left(f_{\phi}(i, u)\right)}{1+\exp \left(f_{\phi}(i, u)\right)},$ $f_{\phi}(i, u)=b_{i}+v_{u}^{T} v_{i}$, and $r$ denote relationships between users and items. $P_{true} (i | u_n, r)$ is the real item distribution for user $u_n$ with relationship $r$, $P_{\theta }\left ( i | u_{n},r \right )$ is the distribution of generated data, and $f_{\phi}\left ( i,u \right )$ indicates the relationship between users and items. Because the sampling of $i$ is discrete, it can be optimized using reinforcement learning based on a policy gradient method \cite{DBLP:journals/ml/Williams92}, instead of the gradient descent method used in the vanilla GAN formulation. However, IRGAN \cite{DBLP:conf/sigir/WangYZGXWZZ17} selects discrete samples from the training data, which causes some intractable problems. The generator of IRGAN produces a separate item ID or an ID list based on the probability calculated by the policy gradient. Under the guidance of the discriminator, the generator may be confused by the items that are simultaneously marked as both "real" and "fake", resulting in a degradation of the performance of the model.  

[\textbf{AdvIR}] Unlike IRGAN, which uses GANs to sample negative documents, AdvIR  \cite{DBLP:conf/www/ParkC19} focuses on selecting more informative negative samples using implicit interactions. This model applies adversarial learning to both the positive and negative interactions in the pair-wise loss function, to capture more informative positive/negative samples. Besides this, the proposed model works for both discrete and continuous inputs. The experimental results were found satisfactory for three tasks relating to ad-hoc retrieval.

[\textbf{CoFiGAN \& ABinCF}] Collaborative filtering GAN (CoFiGAN)  is an extension of IRGAN \cite{DBLP:journals/kbs/LiuPM20}, in which $G$ generates more desirable items using a pair-wise loss, and the $D$ differentiates the generated items from the true ones. CoFiGAN has been experimentally shown to deliver enhanced performance and greater robustness compared to other state-of-the-art models. Another relevant extension of IRGAN is adversarial binary collaborative filtering (ABinCF) \cite{DBLP:conf/aaai/WangSL19}, which adopts the concept of IRGAN for fast RS.

[\textbf{DASO}] Fan et al. \cite{DBLP:conf/ijcai/FanD0WTL19} proposed a GAN-based social recommendation model, referred to as deep adversarial social recommendation (DASO); it uses the minimax game to dynamically guide the informative negative sampling process. For the interactive information, the generator is used to select items based on prior i probabilities and output the user-item pairs as fake samples, whilst the discriminator identifies whether each interaction pair is real or not.

\begin{equation}
\label{eq:eq14}
\begin{aligned}  \min _{\theta_{G}^{I}} \max _{\phi_{D}^{L}} L^{I}(G^{I}, D^{I}) = \sum_{i=1}^{N}(\mathbb{E}_{v \sim p_{real}^{I}(\cdot | u_{i})}[\log D^{I}(u_{i}, v ; \phi_{D}^{I})] + \mathbb{E}_{v \sim G^{I}(\cdot | u_{i} ; \theta_{G}^{I})}[\log (1-D^{I} (u_{i}, v ; \phi_{D}^{I}))]). 
\end{aligned}
\end{equation} 

For social information, the generator is used to select the most relevant friends as informative samples and output fake user-friend pairs, whilst the discriminator identifies the generated user-friend pairs and actual relevant pairs.

\begin{equation}
\label{eq:eq15}
\begin{aligned}  \min _{\theta_{C}^{S}} \max _{\phi_{D}^{S}} L^{S}(G^{S}, D^{S}) = \sum_{i=1}^{N}(\mathbb{E}_{u \sim p_{real}^{S}(\cdot | u_{i})}{[\log D^{S}(u_{i}, v ; \phi_{D}^{S})]} + \mathbb{E}_{u \sim G^{S}(\cdot | u_{i} ; \theta_{G}^{S})}[\log (1-D^{S}(u_{i}, u ; \phi_{D}^{S}))]). \end{aligned} 
\end{equation}

In this way, the user representations are considered using both social and interactive information. In comparisons with other recommendation models \cite{DBLP:conf/cikm/ZhaoMK14, DBLP:conf/sigir/WangYZGXWZZ17}, the authors found that DASO outperforms the DNN-based social recommendation models \cite{DBLP:conf/cikm/ZhaoMK14}. Because of the inefficient calculation of the  softmax function of the generator during gradient descent, the authors replaced softmax with hierarchical softmax, to speed up calculations.

\begin{figure}[pos=htp]
  \centering
    \includegraphics{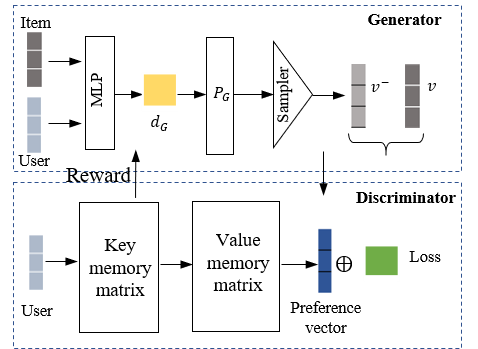}
    
  \centering{ 
      \begin{bfseries}   
          Fig. 5.
      \end{bfseries}
      NMRN-GAN framework \cite{DBLP:conf/kdd/WangYHLWH18}.}
\end{figure}

[\textbf{GAN-HBNR}] Cai et al. \cite{DBLP:conf/aaai/CaiHY18} proposed a GAN based on heterogeneous bibliographic network representation (GAN-HBNR) for citation recommendation; it uses GANs to integrate the heterogeneous bibliographic network structure and vertex content information into a unified framework. It implements a denoising auto-encoder (DAE) \cite{DBLP:conf/icml/VincentLBM08} to generate negative samples, because this produces better representations than a standard auto-encoder. By extracting each continuous vector and concatenating it with the corresponding content vector as the input, GAN-HBNR can learn the optimal representations of content and structure simultaneously, to improve the efficiency of the citation recommendation procedure.

[\textbf{NMRN-GAN}] To capture and store long-term stable interests and short-term dynamic interests, a neural memory streaming GAN (NMRN-GAN) \cite{DBLP:conf/kdd/WangYHLWH18} based on a key-value memory network was proposed for streaming recommendation. It also used the concepts of GANs in negative sampling. More specifically, the generator focused on encouraging the generation of plausible samples to confuse the discriminator. The goal of the discriminator was to separate real items from fake ones produced by the generator. Experiments with optimal hyper-parameters demonstrated that NMRN-GAN significantly outperformed the other comparison models \cite{DBLP:conf/wsdm/WuABSJ17} for two datasets \cite{DBLP:journals/tiis/HarperK16, DBLP:journals/computer/KorenBV09}. However, under the guidance of the discriminator, the generator may become confused by the items simultaneously marked as both "real" and "fake", resulting in a degradation of the performance of the model.

\subsection{Quantitative Analysis}
After introducing the aforementioned GAN-based models designed to mitigate the data noise issue, we summarize and analyze their evaluation metrics, the datasets used in their experiments, and the domains of these datasets. As shown in Table 3, normalized discounted cumulative gain (NDCG) is the most frequently used evaluation metric, and its usage proportion exceeds two-thirds. In terms of the domains of the datasets, the movie dataset is the most popular. Furthermore, we collect the experimental results of several well-established models, for the Movielens dataset, to analyze the differences in their recommendation performance.

\begin{table*}[pos=htp]
\small
\centering{ 
      \begin{bfseries}   
         Table 3
      \end{bfseries}
     Evaluation and domain comparison of GAN-based models for mitigating the data noise issue. (ML: Movielens, PT: Pinterest, GW: Gowalla, AM: Amazon, NF: Netflix, FS: Foursquare, LS: Last.fm, BS: Business, IM: Image, MV: Movie, MS: Music, SO: Social, PP: Paper, BK: Book) \\}
\begin{tabular}{c|l|clllll|l|l}
\hline
\multicolumn{1}{l|}{} &
  \multicolumn{1}{c|}{} &
  \multicolumn{6}{c|}{Evaluation Metric} &
  \multicolumn{1}{c|}{} &
  \multicolumn{1}{c}{} \\ \cline{3-8}
\multicolumn{1}{l|}{\multirow{-2}{*}{Category}} &
  \multicolumn{1}{c|}{\multirow{-2}{*}{Model}} &
  \rotatebox{90}{NDCG} &
  \multicolumn{1}{c}{\rotatebox{90}{HR}} &
  \multicolumn{1}{c}{\rotatebox{90}{Recall}} &
  \multicolumn{1}{c}{\rotatebox{90}{Pre}} &
  \multicolumn{1}{c}{\rotatebox{90}{MAP}} &
  \rotatebox{90}{F1} &
  \multicolumn{1}{c|}{\multirow{-2}{*}{Domain}} &
  \multicolumn{1}{c}{\multirow{-2}{*}{Datasets}} \\ \hline
 &
  APR \cite{DBLP:conf/sigir/0001HDC18} &
  \multicolumn{1}{l}{\checkmark} &
  \checkmark &
   &
   &
   &
   &
  BS, IM &
  Yelp, PT, GW \\
 &
  \cellcolor[HTML]{EFEFEF}AMR \cite{DBLP:journals/corr/abs-1809-07062} &
  \multicolumn{1}{l}{\cellcolor[HTML]{EFEFEF}\checkmark} &
  \cellcolor[HTML]{EFEFEF}\checkmark &
  \cellcolor[HTML]{EFEFEF} &
  \cellcolor[HTML]{EFEFEF} &
  \cellcolor[HTML]{EFEFEF} &
  \cellcolor[HTML]{EFEFEF} &
  \cellcolor[HTML]{EFEFEF}IM &
  \cellcolor[HTML]{EFEFEF}PT, AM \\
 &
  CollaGAN \cite{DBLP:conf/icde/TongLZSC19} &
  \multicolumn{1}{l}{\checkmark} &
   &
   &
  \checkmark &
   &
   &
  MV &
  ML, NF \\
 &
  \cellcolor[HTML]{EFEFEF}\begin{tabular}[c]{@{}l@{}}ACAE/\\ FG-ACAE \cite{DBLP:conf/sigir/YuanYB19}\end{tabular} &
  \multicolumn{1}{l}{\cellcolor[HTML]{EFEFEF}\checkmark} &
  \cellcolor[HTML]{EFEFEF}\checkmark &
  \cellcolor[HTML]{EFEFEF} &
  \cellcolor[HTML]{EFEFEF} &
  \cellcolor[HTML]{EFEFEF} &
  \cellcolor[HTML]{EFEFEF} &
  \cellcolor[HTML]{EFEFEF}MV &
  \cellcolor[HTML]{EFEFEF}ML, Ciao \\
 &
  MASR \cite{DBLP:conf/sigir/TranSL19} &
  \multicolumn{1}{l}{\checkmark} &
  \checkmark &
   &
   &
   &
   &
  MS &
  30MS, AOTM \\
 &
  \cellcolor[HTML]{EFEFEF}SACRA \cite{DBLP:conf/wsdm/LiW020} &
  \multicolumn{1}{l}{\cellcolor[HTML]{EFEFEF}} &
  \cellcolor[HTML]{EFEFEF} &
  \cellcolor[HTML]{EFEFEF} &
  \cellcolor[HTML]{EFEFEF} &
  \cellcolor[HTML]{EFEFEF} \checkmark &
  \cellcolor[HTML]{EFEFEF} &
  \cellcolor[HTML]{EFEFEF}BS &
  \cellcolor[HTML]{EFEFEF}Yelp, FS \\
 &
  FNCF \cite{DBLP:journals/tmm/DuFYXCT19} &
  \multicolumn{1}{l}{\checkmark} &
  \checkmark &
   &
   &
   &
   &
  MV &
  ML \\
\multirow{-8}{*}{\begin{tabular}[c]{@{}c@{}}Models for\\ mitigating\\ casual and\\ malicious\\ noise\end{tabular}} &
  \cellcolor[HTML]{EFEFEF}ATF \cite{DBLP:conf/recsys/ChenL19} &
  \multicolumn{1}{l}{\cellcolor[HTML]{EFEFEF}} &
  \cellcolor[HTML]{EFEFEF} &
  \cellcolor[HTML]{EFEFEF} &
  \cellcolor[HTML]{EFEFEF} &
  \cellcolor[HTML]{EFEFEF} &
  \cellcolor[HTML]{EFEFEF} \checkmark &
  \cellcolor[HTML]{EFEFEF}MV, MS &
  \cellcolor[HTML]{EFEFEF}ML, LS \\ \hline
 &
  IRGAN \cite{DBLP:conf/sigir/WangYZGXWZZ17} &
  \checkmark &
  \multicolumn{1}{c}{} &
  \multicolumn{1}{c}{} &
  \multicolumn{1}{c}{\checkmark} &
  \multicolumn{1}{c}{} &
   &
  MV &
  ML, NF \\
 &
  \cellcolor[HTML]{EFEFEF}DASO \cite{DBLP:conf/ijcai/FanD0WTL19} &
  \cellcolor[HTML]{EFEFEF} \checkmark &
  \multicolumn{1}{c}{\cellcolor[HTML]{EFEFEF}} &
  \multicolumn{1}{c}{\cellcolor[HTML]{EFEFEF}} &
  \multicolumn{1}{c}{\cellcolor[HTML]{EFEFEF} \checkmark} &
  \multicolumn{1}{c}{\cellcolor[HTML]{EFEFEF}} &
  \cellcolor[HTML]{EFEFEF} &
  \cellcolor[HTML]{EFEFEF}SO &
  \cellcolor[HTML]{EFEFEF}Ciao, Epinion \\
 &
  GAN-HBNR \cite{DBLP:conf/aaai/CaiHY18} &
   &
  \multicolumn{1}{c}{} &
  \multicolumn{1}{c}{\checkmark} &
  \multicolumn{1}{c}{} &
  \multicolumn{1}{c}{\checkmark} &
   &
  PP &
  AAN, DBLP \\
 &
  \cellcolor[HTML]{EFEFEF}NMRN-GAN \cite{DBLP:conf/kdd/WangYHLWH18} &
  \cellcolor[HTML]{EFEFEF} \checkmark &
  \multicolumn{1}{c}{\cellcolor[HTML]{EFEFEF} \checkmark} &
  \multicolumn{1}{c}{\cellcolor[HTML]{EFEFEF}} &
  \multicolumn{1}{c}{\cellcolor[HTML]{EFEFEF}} &
  \multicolumn{1}{c}{\cellcolor[HTML]{EFEFEF}} &
  \cellcolor[HTML]{EFEFEF} &
  \cellcolor[HTML]{EFEFEF}MV &
  \cellcolor[HTML]{EFEFEF}ML, NF \\
 &
  ABinCF \cite{DBLP:conf/aaai/WangSL19} &
  \checkmark &
  \multicolumn{1}{c}{} &
  \multicolumn{1}{c}{} &
  \multicolumn{1}{c}{\checkmark} &
  \multicolumn{1}{c}{} &
   &
  MV, BK, BS &
  ML, AM, Yelp \\
 &
  \cellcolor[HTML]{EFEFEF}CoFiGAN \cite{DBLP:journals/kbs/LiuPM20} &
  \cellcolor[HTML]{EFEFEF} \checkmark &
  \multicolumn{1}{c}{\cellcolor[HTML]{EFEFEF}} &
  \multicolumn{1}{c}{\cellcolor[HTML]{EFEFEF}} &
  \multicolumn{1}{c}{\cellcolor[HTML]{EFEFEF} \checkmark} &
  \multicolumn{1}{c}{\cellcolor[HTML]{EFEFEF}} &
  \cellcolor[HTML]{EFEFEF} &
  \cellcolor[HTML]{EFEFEF}MV &
  \cellcolor[HTML]{EFEFEF}ML, NF \\
\multirow{-7}{*}{\begin{tabular}[c]{@{}c@{}}Models for\\ distinguishing\\ the\\ informative\\ samples from\\ unobserved\\ items\end{tabular}} &
  AdvIR \cite{DBLP:conf/www/ParkC19} &
   &
  \multicolumn{1}{c}{} &
  \multicolumn{1}{c}{} &
  \multicolumn{1}{c}{\checkmark} &
  \multicolumn{1}{c}{} &
   &
  MV &
  ML \\ \hline
\end{tabular}
\end{table*}

In the interest of fairness and accuracy, we extracted three studies (FG-ACAE\cite{DBLP:conf/sigir/YuanYB19}, CollaGAN\cite{DBLP:conf/icde/TongLZSC19}, and CoFiGAN\cite{DBLP:journals/kbs/LiuPM20}), all of which conducted experiments on the Movielens-1M dataset. Furthermore, we present their performance comparison in terms of the three most frequently used evaluation metrics. Table 4 shows that IRGAN is the most commonly used baseline model, and most of the baselines are based on neural networks. Specifically, FG-ACAE consistently outperforms ACAE, owing to the increased quantity of adversarial noise. Both FG-ACAE and ACAE outperform APR (the first adversarial learning model) on the three chosen metrics. CGAN outperforms NeuMF, which indicates that adversarial training can boost the accuracy of the recommendation models. However, NeuMF performs significantly better than IRGAN, demonstrating that neural networks can extract better latent features than conventional models based on matrix factorization (MF). A similar pattern also appears in CoFiGAN's experiment. To summarize, adversarial training can improve model performances, especially in neural network-based models.

\begin{table}[pos=htp]

\centering{ 
      \begin{bfseries}   
         Table 4
      \end{bfseries}
     Performance comparison of GAN-based models for mitigating the data noise problem, tested on the Movielens-1M datasets (results for FG-ACAE \cite{DBLP:conf/sigir/YuanYB19}, CollaGAN \cite{DBLP:conf/icde/TongLZSC19}, and CoFiGAN \cite{DBLP:journals/kbs/LiuPM20}).}

\begin{tabular}{c|c|ccllll}
\hline
 &
   &
  \multicolumn{6}{c|}{Evaluation Metric} \\ \cline{3-8} 
\multirow{-2}{*}{Source} &
  \multirow{-2}{*}{Model} &
  \multicolumn{1}{l}{HR@5} &
  \multicolumn{1}{l}{HR@10} &
  Pre@5 &
  Pre@10 &
  NDCG@5 &
  NDCG@10 \\ \hline
 &
  NeuMF &
  \multicolumn{1}{l}{0.5832} &
  \multicolumn{1}{l}{0.7250} &
  \multicolumn{1}{c}{/} &
  \multicolumn{1}{c}{/} &
  0.4304 &
  0.4727 \\
 &
  \cellcolor[HTML]{EFEFEF}APR &
  \multicolumn{1}{l}{\cellcolor[HTML]{EFEFEF}0.5875} &
  \multicolumn{1}{l}{\cellcolor[HTML]{EFEFEF}0.7263} &
  \multicolumn{1}{c}{\cellcolor[HTML]{EFEFEF}/} &
  \multicolumn{1}{c}{\cellcolor[HTML]{EFEFEF}/} &
  \cellcolor[HTML]{EFEFEF}0.4314 &
  \cellcolor[HTML]{EFEFEF}0.4763 \\
 &
  ACAE &
  \multicolumn{1}{l}{0.5988} &
  \multicolumn{1}{l}{0.7379} &
  \multicolumn{1}{c}{/} &
  \multicolumn{1}{c}{/} &
  0.4446 &
  0.4905 \\
\multirow{-4}{*}{\begin{tabular}[c]{@{}c@{}}FG-ACAE\\ \cite{DBLP:conf/sigir/YuanYB19} \end{tabular}} &
  \cellcolor[HTML]{EFEFEF}FG-ACAE &
  \multicolumn{1}{l}{\cellcolor[HTML]{EFEFEF}0.6186} &
  \multicolumn{1}{l}{\cellcolor[HTML]{EFEFEF}0.7507} &
  \multicolumn{1}{c}{\cellcolor[HTML]{EFEFEF}/} &
  \multicolumn{1}{c}{\cellcolor[HTML]{EFEFEF}/} &
  \cellcolor[HTML]{EFEFEF}0.4586 &
  \cellcolor[HTML]{EFEFEF}0.5136 \\ \hline
 &
  NeuMF &
  / &
  / &
  0.394 &
  0.363 &
  0.402 &
  0.447 \\
 &
  \cellcolor[HTML]{EFEFEF}IRGAN &
  \cellcolor[HTML]{EFEFEF}/ &
  \cellcolor[HTML]{EFEFEF}/ &
  \cellcolor[HTML]{EFEFEF}0.375 &
  \cellcolor[HTML]{EFEFEF}0.351 &
  \cellcolor[HTML]{EFEFEF}0.382 &
  \cellcolor[HTML]{EFEFEF}0.415 \\
\multirow{-3}{*}{\begin{tabular}[c]{@{}c@{}}CollaGAN\\ \cite{DBLP:conf/icde/TongLZSC19} \end{tabular}} &
  CollaGAN &
  / &
  / &
  0.428 &
  0.398 &
  0.417 &
  0.458 \\ \hline
 &
  \cellcolor[HTML]{EFEFEF}NeuMF &
  \cellcolor[HTML]{EFEFEF}/ &
  \cellcolor[HTML]{EFEFEF}/ &
  \cellcolor[HTML]{EFEFEF}0.3993 &
  \cellcolor[HTML]{EFEFEF}0.3584 &
  \cellcolor[HTML]{EFEFEF}0.4133 &
  \cellcolor[HTML]{EFEFEF}0.3844 \\
 &
  IRGAN &
  / &
  / &
  0.3098 &
  0.2927 &
  0.3159 &
  0.3047 \\
\multirow{-3}{*}{\begin{tabular}[c]{@{}c@{}}CoFiGAN\\ \cite{DBLP:journals/kbs/LiuPM20} \end{tabular}} &
  \cellcolor[HTML]{EFEFEF}CoFiGAN &
  \cellcolor[HTML]{EFEFEF}/ &
  \cellcolor[HTML]{EFEFEF}/ &
  \cellcolor[HTML]{EFEFEF}0.4484 &
  \cellcolor[HTML]{EFEFEF}0.4007 &
  \cellcolor[HTML]{EFEFEF}0.4641 &
  \cellcolor[HTML]{EFEFEF}0.4300 \\ \hline
\end{tabular}
\end{table}

\subsection{Qualitative Analysis}
To more clearly elucidate the specific designs of the aforementioned models, we demonstrate their particular designs and analyze their advantages, as shown in Table 5.

\begin{table*}[pos=htp]

\centering{ 
      \begin{bfseries}   
         Table 5
      \end{bfseries}
     A schematic representation of GAN-based RS for mitigating the data noise issue.}

\begin{tabular}{|c|c|l|l|l|}
\hline
Category &
  Model &
  \multicolumn{1}{c|}{Generator} &
  \multicolumn{1}{c|}{Discriminator} &
  \multicolumn{1}{c|}{Advantage} \\ \hline
\multirow{6}{*}{\begin{tabular}[c]{@{}c@{}}Models\\ for\\ mitigating\\ casual and\\ malicious\\ noise\end{tabular}} &
  \begin{tabular}[c]{@{}c@{}}APR\\ \cite{DBLP:conf/sigir/0001HDC18} \end{tabular} &
  \begin{tabular}[c]{@{}l@{}}Maximize the BPR loss \\ function to learn\\ adversarial noise\end{tabular} &
  \begin{tabular}[c]{@{}l@{}}Minimize the BPR loss\\ function to improve\\ model robustness \end{tabular} &
  \begin{tabular}[c]{@{}l@{}}Add adversarial \\ perturbations to the\\ model parameters\end{tabular} \\ \cline{2-5} 
 &
  \begin{tabular}[c]{@{}c@{}}AMR\\ \cite{DBLP:journals/corr/abs-1809-07062} \end{tabular} &
  \begin{tabular}[c]{@{}l@{}}Maximize the VBPR \\ loss function\end{tabular} &
  \begin{tabular}[c]{@{}l@{}}Minimize the VBPR loss\\ function to improve \\ model robustness \end{tabular} &
  \begin{tabular}[c]{@{}l@{}}Apply adversarial \\ noise to the image \\ RS\end{tabular} \\ \cline{2-5} 
 &
  \begin{tabular}[c]{@{}c@{}}MASR\\ \cite{DBLP:conf/sigir/TranSL19} \end{tabular} &
  \begin{tabular}[c]{@{}l@{}}Maximize the BPR loss\\ function\end{tabular} &
  \begin{tabular}[c]{@{}l@{}}Minimize the BPR loss \\ function to improve \\ model robustness \end{tabular} &
  \begin{tabular}[c]{@{}l@{}}Use APR as \\ part of the music \\ sequence \\ recommendation model\end{tabular} \\ \cline{2-5} 
 &
  \begin{tabular}[c]{@{}c@{}}CollaGAN\\ \cite{DBLP:conf/icde/TongLZSC19} \end{tabular} &
  \begin{tabular}[c]{@{}l@{}}Apply variational\\ autoencoder as the\\ generator\end{tabular} &
  \begin{tabular}[c]{@{}l@{}}Maximize the\\ probability of generated\\ item samples\end{tabular} &
  \begin{tabular}[c]{@{}l@{}}The first method\\ that applies GANs to \\ collaborative filtering\end{tabular} \\ \cline{2-5} 
 &
  \begin{tabular}[c]{@{}c@{}}ACAE\\ \cite{DBLP:conf/ijcnn/YuanYB19} \end{tabular} &
  \begin{tabular}[c]{@{}l@{}}Add adversarial noise\\ at different locations\\ in the model\end{tabular} &
  \begin{tabular}[c]{@{}l@{}}Identify the\\ rating vectors\end{tabular} &
  \begin{tabular}[c]{@{}l@{}}Indicate how to\\ balance the accuracy\\ and robustness\end{tabular} \\ \cline{2-5} 
 &
  \begin{tabular}[c]{@{}c@{}}ATF\\ \cite{DBLP:conf/recsys/ChenL19} \end{tabular} &
  \begin{tabular}[c]{@{}l@{}}Maximize the BPR loss\\ function\end{tabular} &
  \begin{tabular}[c]{@{}l@{}}Minimize the BPR loss\\ function\end{tabular} &
  \begin{tabular}[c]{@{}l@{}}The first method to \\ combine tensor \\ factorization and GANs\end{tabular} \\ \hline
\multirow{5}{*}{\begin{tabular}[c]{@{}c@{}}Models\\ for\\ distinguishing\\ the\\ uninformative\\ samples\\ from\\ unobserved\\ items\end{tabular}} &
  \begin{tabular}[c]{@{}c@{}}IRGAN\\ \cite{DBLP:conf/sigir/WangYZGXWZZ17} \end{tabular} &
  \begin{tabular}[c]{@{}l@{}}Select items from\\ the set of existing\\ items\end{tabular} &
  \begin{tabular}[c]{@{}l@{}}Determine the \\ relationship pairs \\ as real or generated\end{tabular} &
  \begin{tabular}[c]{@{}l@{}}The first method to use \\ GANs in informative\\ retrieval\end{tabular} \\ \cline{2-5} 
 &
  \begin{tabular}[c]{@{}c@{}}CoFiGAN\\ \cite{DBLP:journals/kbs/LiuPM20} \end{tabular} &
  \begin{tabular}[c]{@{}l@{}}Minimize the distance\\ between generated \\ and real samples\end{tabular} &
  \begin{tabular}[c]{@{}l@{}}Like the discriminator\\ in IRGAN\end{tabular} &
  An extension of IRGAN \\ \cline{2-5} 
 &
  \begin{tabular}[c]{@{}c@{}}DASO\\ \cite{DBLP:conf/ijcai/FanD0WTL19} \end{tabular} &
  \begin{tabular}[c]{@{}l@{}}Apply two generators\\ for social and \\ interactive information\end{tabular} &
  \begin{tabular}[c]{@{}l@{}}Identify whether \\ each interaction\\ pair is real\end{tabular} &
  \begin{tabular}[c]{@{}l@{}}Generate valuable\\ negative samples to learn\\ better representations\end{tabular} \\ \cline{2-5} 
 &
  \begin{tabular}[c]{@{}c@{}}GAN-HBNR\\ \cite{DBLP:conf/aaai/CaiHY18} \end{tabular} &
  \begin{tabular}[c]{@{}l@{}}Apply DAE to integrate\\ the content and \\ structure of the \\ heterogeneous network\end{tabular} &
  An energy function &
  \begin{tabular}[c]{@{}l@{}}Integrate the network \\ structure and vertex \\ content into a\\ unified framework\end{tabular} \\ \cline{2-5} 
 &
  \begin{tabular}[c]{@{}c@{}}NMRN-GAN\\ \cite{DBLP:conf/kdd/WangYHLWH18} \end{tabular} &
  \begin{tabular}[c]{@{}l@{}}Generate more \\ recognizable\\ negative samples\end{tabular} &
  \begin{tabular}[c]{@{}l@{}}Identify negative \\ samples from the \\ real data\end{tabular} &
  \begin{tabular}[c]{@{}l@{}}Apply GANs in the process\\ of negative sampling for\\ stream recommendation\\ model\end{tabular} \\ \hline
\end{tabular}
\end{table*}

\section{GAN-Based Recommendation Models for Mitigating the Data Sparsity Issue}
Alongside data noise, data sparsity is another severe problem in RSs. In this section, we highlight some typical models, to identify the most notable and promising advancements of recent years. We divide them into two categories: (1) models for generating user preferences by augmenting interactive information, and (2) models for synthesizing user preferences by augmenting them with auxiliary information.

\subsection{Models for Generating User Preferences by Augmenting with Interactive Information}
Several productive approaches have enabled GAN-based architectures to improve the utility of RSs, by augmenting them with missing interactive information and thereby mitigating the data sparsity issue; these include conditional filtered GAN (CFGAN)  \cite{DBLP:conf/cikm/ChaeKKL18}, AugCF  \cite{DBLP:conf/kdd/WangYWNHC19}, PLASTIC  \cite{DBLP:conf/ijcai/ZhaoWYGYC18}, and more.

\begin{figure}[pos=htp]
  \centering
    \includegraphics{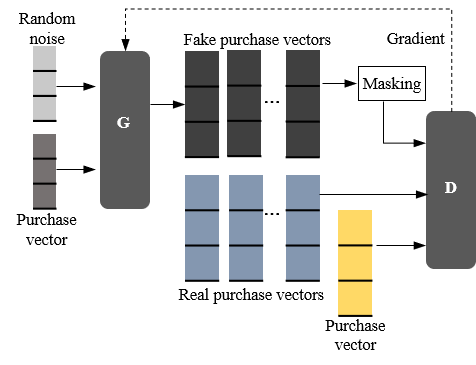}
    
  \centering{ 
      \begin{bfseries}   
          Fig. 6.
      \end{bfseries}
      CFGAN framework \cite{DBLP:conf/cikm/ChaeKKL18}.}
\end{figure}

[\textbf{CFGAN \& RAGAN}] CFGAN \cite{DBLP:conf/cikm/ChaeKKL18} was the first model to generate the purchase vectors of users rather than item IDs; it was inspired by the CGAN concept \cite{DBLP:journals/corr/MirzaO14}. The framework of CFGAN is shown in Fig. 6. The loss function of $G$ is

\begin{equation}
\label{eq:eq16}
L^{G}=\sum_{u} \log (1-(D(\widehat{r}_{u} \cdot e_{u}) | c_{u})),
\end{equation}where the input of $G$ is the combination of a purchase vector $c_u$ for user $u$ and the random noise $z$. The $G$ generates a low-dimensional user preference vector $\hat{r_{u}}$ using a multi-layer neural network. The $D$ distinguishes generated preference vectors from real purchase vectors. Its loss function is

\begin{equation}
\label{eq:eq17}
\begin{aligned}
L^{D} &=-\mathbb{E}_{x \sim P_{data}}[\log D(x | c)]-\mathbb{E}_{\hat{x} \sim P_{\phi }}[\log (1-D(\hat{x} | c))]  \\& =-\sum_{u}(\log D(r_{u} | c_{u})+\log (1-D((\widehat{r_{u}} \cdot e_{u}) | c_{u}))),
\end{aligned}
\end{equation}where $P_{data}$ represents the real data distribution, $P_{\phi}$ is the data distribution generated by the generator, $\cdot$ represents the multiplication of the elements, and $e_u$ is an indicator vector specifying whether or not $u$ has purchased item $i$. To better simulate the preferences of users, this model uses $e_{u}$ as the masking mechanism. 

In terms of accuracy, CFGAN has outperformed other state-of-the-art models (including IRGAN \cite{DBLP:conf/sigir/WangYZGXWZZ17} and GraphGAN \cite{DBLP:conf/aaai/WangWWZZZXG18}) by at least a 2.8\% enhancement for three different datasets: Ciao \cite{DBLP:conf/wsdm/TangGL12}, Watcha \cite{DBLP:conf/naacl/BaldwinC12}, and Movielens \cite{DBLP:journals/tiis/HarperK16}. This is a new direction in vector-wise adversarial training for GAN-based recommendation tasks; it prevents $D$ from being misled. Several state-of-the-art GANs have achieved better stability than CFGAN and can be applied thereto, to further improve the recommendation accuracy. Besides this, Chae et al. \cite{DBLP:conf/www/ChaeKKC19} proposed a rating augmentation model based on GANs, referred to as RAGAN. It uses the observed data to learn initial parameters and then generate plausible data via its generator. Finally, the augmented data are used to train conventional conditional filtering (CF) models.

[\textbf{UGAN}] Inspired by the advances of CFGAN (in terms of generating vectors instead of item lists), Wang et al. \cite{DBLP:conf/pakdd/WangGWYWX19} proposed a unified GAN (UGAN) to alleviate the data sparsity problem. The main objective of the UGAN is to generate user profiles with rating information, by capturing the input data distribution. The discriminator uses the Wasserstein distance to distinguish the generated samples from the real ones. The authors evaluated its recommendation performance on two public datasets. The experimental results show that the UGAN achieves significant improvement.

[\textbf{AugCF}] AugCF \cite{DBLP:conf/kdd/WangYWNHC19} is another GAN-based CF model designed to generate interactive information. It generates interactions for different recommendation tasks under different auxiliary information conditions. It features two training stages: (1) The generator generates the most preferred item for the user in the interaction category. The generated tuple (user, item, and interaction category) can be considered as a valid and realistic sample of the original dataset. The discriminator is only used to determine whether the generated data tuple is real. (2) The generator is fixed and used only to generate data. Then, the discriminator is used to determine whether the user likes the items or not.

\begin{figure}[pos=htp]
  \centering
    \includegraphics{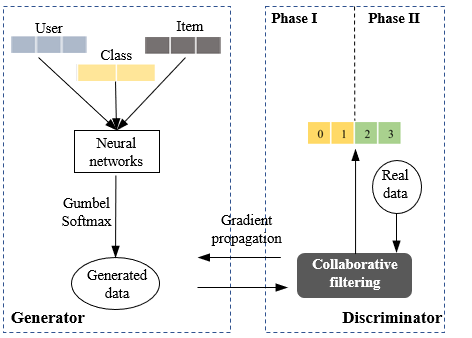}
    
  \centering{ 
      \begin{bfseries}   
          Fig. 7.
      \end{bfseries}
      AugCF framework \cite{DBLP:conf/kdd/WangYWNHC19}.}
\end{figure}

The loss function of AugCF is defined as Eq.\ref{eq:eq18}. The discriminator and generator compete on the category label $c$ and user $u$. To perform the different roles in the two phases of the discriminator model, AugCF expands the first two relationship categories (like or dislike) into four ones: true \& like, true \& dislike, false \& like, and false \& dislike. Its loss function is 

\begin{equation}
\label{eq:eq18}
\begin{aligned}L^{G, D}=\min _{\theta} \max _{\phi}(\mathbb{E}_{(u, v, y) \sim P_{C}(v|u, c)} \log [D_{\phi}(v, y | u, c) ] + \mathbb{E}_{(u, v, y) \sim P_{G_{\theta}}(v|u, c)} \log [D_{\phi}(v, y | u, c) ]),
\end{aligned}
\end{equation}
where $P_{G_{\theta}}\left ( v|u,c \right )$ and $P_{c}\left ( v|u,c \right )$ represent the distributions of the generated and real data, respectively. The models were experimentally investigated using user reviews as auxiliary information; the results show that AugCF outperformed the baselines and state-of-the-art models. The AugCF provides a new method for alleviating the data sparsity problem; however, this problem remains a long-standing research topic in the field of RSs.

\begin{figure}[pos=htp]
  \centering
    \includegraphics{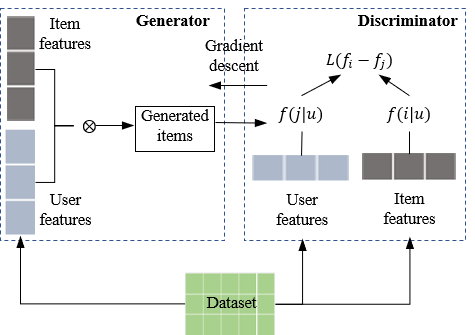}
    
  \centering{ 
      \begin{bfseries}   
          Fig. 8.
      \end{bfseries}
      APL framework \cite{DBLP:journals/eswa/SunWWY19}.}
\end{figure}

[\textbf{APL}] Besides this, Sun et al. \cite{DBLP:journals/eswa/SunWWY19} proposed advanced pairwise learning (APL), which is a general GAN-based pair-wise learning framework. APL combines the generator and discriminator via adversarial pair-wise learning, based on the assumption that users prefer items that have already been consumed. Under this framework, the generator $g_{\theta}$ attempts to generate items that approximate the real distribution for each user. The discriminator $f_{\phi}$ learns the ranking function between two pairs of items and determines the preference of each user. Expressed otherwise, for each pair of items $i$ and $j$, the discriminator must identify which is more in line with user preferences. APL directly uses pair-wise ranking as the loss function, instead of a function based on the probability distribution; The loss function of APL is defined as follows:

\begin{equation}
\label{eq:eq19}
L\left(g_{\theta}, f_{\phi}\right)=\max _{\theta} \min _{\phi} \sum_{u=1}^{m} \mathbb{E}_{i \sim P_{real}(i | u)}^{ j \sim P_{\theta}(j | u)} L\left(f_{\phi}(i | u)-f_{\phi}(j | u)\right).
\end{equation}

Here, $L(x)$ is the pair-wise ranking loss function, which differs from the loss function of GANs. If the discriminant loss function is specifically designed to maximize the difference in ranking scores between the observed and generated items, then the original objective function is equivalent to that of WGAN \cite{DBLP:conf/iclr/ArjovskyB17}, as shown in Eq.\ref{eq:eq20}.

\begin{equation}
\label{eq:eq20}
\begin{aligned}
L\left(g_{\theta}, f_{\phi}\right) & =\max _{\theta} \min _{\phi} \sum_{u=1}^{m}-\mathbb{E}_{i \sim P_{real}(i|u)}^{j \sim P_{\theta}(j|u)} L\left(f_{\phi}(i|u)-f_{\phi}(j|u)\right) \\& =\min _{\theta} \max _{\phi} \sum_{u=1}^{m} [\mathbb{E}_{i \sim P_{\theta}(i|u)} f_{\phi}(i|u) - \mathbb{E}_{j \sim P_{\theta}(j|u)} f_{\phi}(j|u)].
\end{aligned}
\end{equation}

This model manages the problem of gradient vanishing by utilizing the pair-wise loss function and the Gumbel--Softmax technique \cite{DBLP:conf/iclr/JangGP17}. Extensive experiments have demonstrated its effectiveness and stability. APL is a general adversarial learning recommendation framework that can be used for future RSs in various fields.

[\textbf{PLASTIC}] GANs have also been applied in numerous other recommendation fields, to generate interactions and mitigate data sparsity. For example, PLASTIC \cite{DBLP:conf/ijcai/ZhaoWYGYC18} was proposed for sequence recommendation; it combines MF, recurrent neural networks (RNNs), and GANs. Its framework is shown in Fig. 9.

\begin{figure}[pos=htp]
  \centering
    \includegraphics{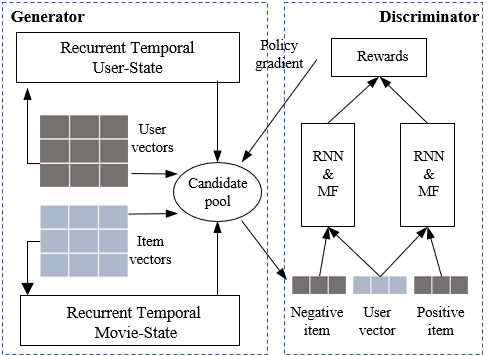}
    
  \centering{ 
      \begin{bfseries}   
          Fig. 9.
      \end{bfseries}
      PLASTIC framework \cite{DBLP:conf/ijcai/ZhaoWYGYC18}.}
\end{figure}

In the adversarial training process, the generator---like that of the CGAN \cite{DBLP:journals/corr/MirzaO14}---takes users and times as inputs, to directly predict the recommendation list of the user. For the discriminator, PLASTIC integrates long-term and short-term ranking models through a Siamese network, to correctly distinguish real samples from generated ones. Extensive experiments have shown that it achieves better performance than other models \cite{DBLP:conf/wsdm/WuABSJ17, DBLP:conf/sigir/WangYZGXWZZ17}. In addition to PLASTIC, Yoo et al. \cite{DBLP:journals/corr/YooHYRKH0Y17} combined an energy based GAN and sequence GAN, to learn the sequential preferences of users and predict the next recommended items. 

[\textbf{RecGAN}] Recurrent GAN (RecGAN) combines recurrent recommender networks and GANs to learn temporal features of users and items \cite{DBLP:conf/recsys/BharadhwajPL18}. In RecGAN, the generator predicts a sequence of items for a user, by fitting the distribution of items. The discriminator determines whether the sampled items are from the distribution of the user's real preferences. In experiments, RecGAN outperformed all baseline models---including recurrent recommender networks \cite{DBLP:conf/wsdm/WuABSJ17} and AutoRec  \cite{DBLP:conf/iconip/OuyangLRX14}---on movie and food recommendation datasets \cite{DBLP:journals/tiis/HarperK16, DBLP:journals/computer/KorenBV09}. This work lacks an in-depth study of the feasibility and impacts of gated recurrent unit (GRU)  modifications in reflecting different granularities of temporal patterns.

[\textbf{APOIR}] Zhou et al. \cite{DBLP:conf/www/0002YZTZW19} proposed adversarial point-of-interest recommendation (APOIR), to learn the potential preferences of users in point-of-interest (POI) recommendations. The generator selects a set of POIs using the policy gradient and tries to match the real distribution. Then, the discriminator distinguishes the generated POIs from the actual browsing behaviors of the user. The proposed model uses geographical features and the social relationships of users to optimize the generator. The loss function of APOIR is

\begin{equation}
\label{eq:eq21}
\begin{aligned}
L^{G, D}= \min _{\theta} \max _{\phi} \sum_{u_{i}}(\mathbb{E}_{l^{+}}[\log D_{\phi}(u_{i}, l^{+})] + \mathbb{E}_{l^{R} \sim R_{\theta}(l^{R} | u_{i})}[\log (1-D_{\phi}(u_{i}, l^{R}))]),
\end{aligned}
\end{equation}where $l^{+}$ represents the POIs already visited, and $D_{\phi}(u_i, l^{R})$ evaluates the probability that user $u_i$ has preferentially visited POI $l^R$. Once the confrontation between the generator and discriminator has been balanced, the recommender (generator) $R_{\theta} (l^R |u_i)$ recommends high-quality POIs for the user.

[\textbf{Geo-ALM}] Liu et al. \cite{DBLP:conf/ijcai/0061W0Y19} proposed a geographical information-based adversarial learning model (Geo-ALM), to combine geographical information and GANs for making POI recommendations. In the model, the $G$ generates unvisited POIs that match user preferences as much as possible, and the $D$ focuses on distinguishing the visited POIs from unvisited ones as accurately as possible. The authors verified the superiority of Geo-ALM on two public datasets: Foursquare and Gowalla.

\subsection{Models for Synthesizing User Preferences by Augmenting with Auxiliary Information}
In addition to augmenting models with interactive information to directly generate user preferences, several studies have tried to use the generators of GAN-based architectures to augment models with auxiliary information \cite{DBLP:conf/www/ChaeKKC19, DBLP:journals/corr/abs-1909-03529, DBLP:conf/icdm/YangGWJXW18, DBLP:conf/www/PereraZ19, DBLP:journals/corr/abs-1903-10794}.

[\textbf{RSGAN}] Region-separative GAN (RSGAN) \cite{DBLP:journals/corr/abs-1909-03529} was proposed to augment social recommenders with more reliable friends and alleviate the problem of data sparsity. It consists of two components: $G$ and $D$. $G$ is responsible for generating reliable friends and the items consumed by these friends. The model first constructs a heterogeneous network, to identify seed friends with higher reliability. The model collects seed users for each user and encodes them into binary vectors as the incomplete social preferences of the user. Then, the probability distributions of friends with high likelihoods are sampled through Gumbel--Softmax \cite{DBLP:conf/iclr/JangGP17}. This strategy is also used to simulate the sampling of items. To order the candidate items, RSGAN adopts the idea of social BPR \cite{DBLP:conf/cikm/ZhaoMK14} in its $D$. It sorts the candidates and recommends an item list for each user. If the items consumed by the generated friends are not useful, the $D$ punishes them and returns the gradient to the $G$, to reduce the likelihood of generating such friends. The loss function of RSGAN is

\begin{equation}
\label{eq:eq22}
\begin{aligned} L^{D, G} =\min _{D} \max _{G}-\mathbb{E}((\log \sigma(x_{ui}-x_{uz}) +\log \sigma(x_{uz}-x_{uj}))). 
\end{aligned}
\end{equation}

The designers of RSGAN conducted experiments with three models: conventional social recommendation models \cite{DBLP:conf/aaai/HeM16, DBLP:conf/cikm/ZhaoMK14}, DNN-based models \cite{DBLP:conf/www/HeLZNHC17}, and other GAN-based ones \cite{DBLP:conf/sigir/WangYZGXWZZ17, DBLP:conf/cikm/ChaeKKL18}. The experimental results show that RSGAN outperforms all other methods in ranking prediction. The possible reason for this is that RSGAN builds a dynamic framework to generate friend relationships and thereby alleviate the data sparsity problem.

\begin{figure}[pos=htp]
  \centering
    \includegraphics{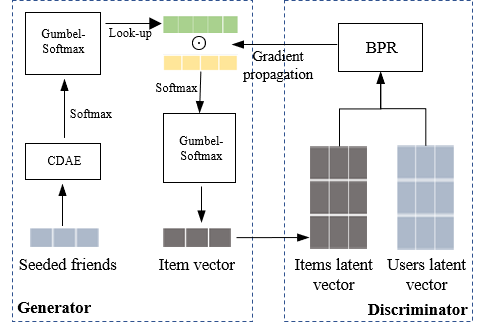}
    
  \centering{ 
      \begin{bfseries}   
          Fig. 10.
      \end{bfseries}
      RSGAN framework \cite{DBLP:journals/corr/abs-1909-03529}.}
\end{figure}

[\textbf{KTGAN}] KTGAN   \cite{DBLP:conf/icdm/YangGWJXW18} was proposed to augment data and alleviate the problem of data sparsity by importing external information. The model consists of two phases: (1) extracting feature embeddings using auxiliary and interactions information, to construct the initial representations of users and items; and (2) putting these vectors into an IRGAN-based generator and discriminator for adversarial learning. The discriminator attempts to identify whether the user-item pair is generated or real. The loss function of KTGAN is

\begin{equation}
\label{eq:eq23}
\begin{aligned}
L^{D, G} = \min _{D} \max _{G} \sum_{n=1}^{N}(\mathbb{E}_{m \sim p_{true}(m|u_{n}, r)}[\log P(m | u_{n})] + \mathbb{E}_{m \sim p_{\theta}(m|u_{n}, r)}[\log (1-P(m|u_{n}))]),
\end{aligned}
\end{equation}
where $P(m|u_n )$ estimates the probability that user $u_n$ prioritizes item $m$. The parameter $r$ represents the relationship between a user and an item. The experiments show that it achieves a better accuracy and NDCG than other methods \cite{DBLP:conf/www/HeLZNHC17, DBLP:conf/sigir/WangYZGXWZZ17, DBLP:conf/aaai/WangWWZZZXG18}. In particular, if the generator can generate a more accurate low-dimensional vector at the start of training, it can train a better-optimized discriminator and improve the performance of the model.

\begin{figure}[pos=htp]
  \centering
    \includegraphics{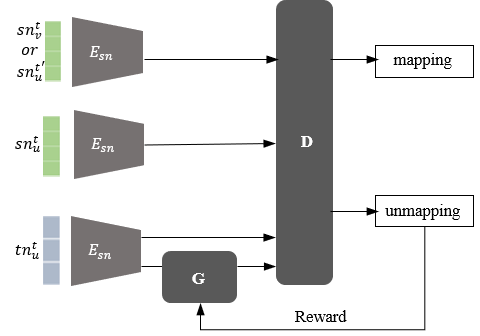}
    
  \centering{ 
      \begin{bfseries}   
          Fig. 11.
      \end{bfseries}
      CnGAN framework \cite{DBLP:conf/www/PereraZ19}.}
\end{figure}

[\textbf{CnGAN}] Perera et al. \cite{DBLP:conf/www/PereraZ19} proposed a cross-network  GAN (CnGAN), to learn the mapping encoding between the target and source domains for non-overlapped users in different domains. The framework is shown in Fig. 11. $E$ is the neural network encoder that converts the distribution of inputs into a vector; the generator $G$ uses the target encoding as an input, to generate the mapping encoding $E_{sn}$ of the source domain for non-overlapping users. Moreover, $G$ makes the generated preference domain correspond as much as possible to the real source domain, to deceive the discriminator. The loss function is

\begin{equation}
\label{eq:eq24}
\begin{aligned} \min _{G} L^{G}=\mathbb{E}_{tn_{u}^{t} \sim p_{data}(tn)} L_{fake}\left(E_{tn}\left(tn_{u}^{t}\right), G\left(E_{tn}\left(tn_{u}^{t}\right)\right)\right) + \mathbb{E}_{tn_{u}^{t}, sn_{u}^{t} \sim p_{data}(tn, sn)} L_{content}\left(E_{sn}\left(sn_{u}^{t}\right), G\left(E_{tn}\left(tn_{u}^{t}\right)\right)\right).
\end{aligned}
\end{equation}

$D$ distinguishes the real source-domain encoding from the generated one. In particular, the mismatching source- and target-domain encodings are used as the input of $G$. The overlap between the user's real target- and source-domain embeddings is taken as the actual mapping. Formally, the loss function of $D$ is

\begin{equation}
\label{eq:eq25}
\begin{array}{l}{\max _{E_{tn}, E_{s n}, D} L(E_{t n}, E_{s n}, D)} \\ {=\mathbb{E}_{t n_{u}^{t}, s n_{u}^{t} \sim p_{data}(t n, s n)} L_{r e a l}(E_{t n}(t n_{u}^{t}), E_{s n}(s n_{u}^{t}))} \\ +\mathbb{E}_{tn_{u}^{t} \sim p_{data}(tn)} L_{fake}(E_{t n}(t n_{u}^{t}), G(E_{t n}(t n_{u}^{t}))) \\ {+\mathbb{E}_{t n_{u}^{t}, \overline{s n}_{u}^{t} \sim \bar{p}_{data}(t n, \overline{s n})} L_{ mismatch}(E_{t n}(t n_{u}^{t}), E_{s n}(\overline{sn}_{u}^{t}))},\end{array}
\end{equation}
where $P_{data}\left ( tn,sn \right )$ is a matching pair with a mapping relationship, $\overline{P}_{data}\left ( tn,\overline{sn} \right )$ is a matching pair with no mapping relationship, $P_{data}\left ( tn \right )$ is the local distribution of the target domain, and $G(x)$ is the matching source-domain encoding, generated for a given target domain. CnGAN employs a new approach to alleviating the data sparsity issue. It represents the first attempt to use GANs to generate missing source-domain preferences for non-overlapping users, by generating mapping relationships between the source and target domains.

[\textbf{RecSys-DAN \& AB-GAN}] RecSys-DAN , proposed by Wang et al. \cite{DBLP:journals/corr/abs-1903-10794}, is similar to CnGAN \cite{DBLP:conf/www/PereraZ19}. It also uses an adversarial approach to transfer the potential representations of users and items from different domains to the target domain. Besides this, Zhang et al. \cite{DBLP:conf/mm/ZhengSCHCN19} designed and implemented a virtual try on GAN model---referred to as AB-GAN ---in which the generator $G$ generates images for the modeling of 2D images. Given four features (the user-image feature, desired-posture feature, new-clothing feature, and body shape mask), an image of the user wearing new clothing is generated. The $G$ can synthesize an original image of a person and ensure that the data distribution is similar to that of the real image. AB-GAN is superior to other advanced methods, according to qualitative analysis and quantitative experiments.

[\textbf{ATR \& DVBPR}] In the adversarial training for review-based recommendation (ATR) model, Rafailidis et al. \cite{DBLP:conf/sigir/RafailidisC19} used GANs to generate reviews likely to be relevant to the user's preferences. The discriminator focuses on distinguishing between the generated reviews and those written by users. Similar to generating user-related reviews, the generator also generates items-related reviews. After obtaining review information through adversarial learning, this model predicts user preferences through the joint factorization of rating information. Furthermore, to synthesize images that are highly consistent with users' preferences in fashion recommendation, Kang et al. \cite{DBLP:conf/icdm/KangFWM17} proposed deep visually aware BPR (DVBPR), in which the generator generates appropriate images that look realistic, and the discriminator tries to distinguish generated images from the real ones using a Siamese-CNN framework. DVBPR is the first model to exploit the generative power of GANs in fashion recommendations. The same ideas can also be applied to non-visual content.

[\textbf{LARA}] Sun et al. \cite{DBLP:conf/wsdm/SunLLRGN20} proposed the end-to-end adversarial framework LARA ; it uses multiple generators to generate user profiles from various item attributes. In LARA, every single attribute of each item is input into a unique generator. The outputs of all generators are integrated through a neural network, to obtain the final representation vector. The discriminator distinguishes the real item-user pair from the three interaction pairs, which include the item-generated user, item-interested real user, and item-uninterested real user. The architecture of LARA is shown in Fig. 12. Formally, the objective function of LARA is

\begin{equation}
\label{eq:eq26}
\begin{aligned}
\mathcal{L}^{G^{*}, D^{*}}=\min _{\theta} \max _{\phi} \sum_{n=1}^{N}\left(\mathbb{E}_{\mathbf{u}^{+} \sim p_{t r u e}\left(\mathbf{u}^{+} | I_{n}\right)}\left[\log \left(D\left(\mathbf{u}^{+} | I_{n}\right)\right)\right]\right.\\
+\mathbb{E}_{\mathbf{u}^{c} \sim p_{\theta}\left(\mathbf{u}^{c} | I_{n}\right)}\left[\log \left(1-D\left(\mathbf{u}^{c} | I_{n}\right)\right)\right] \\
+\mathbb{E}_{\mathbf{u}^{-} \sim p_{f a l s e}\left(\mathbf{u}^{-} | I_{n}\right)}\left[\log \left(1-D\left(\mathbf{u}^{-} | I_{n}\right)\right)\right]).
\end{aligned}
\end{equation}

\begin{figure}[pos=htp]
  \centering
    \includegraphics{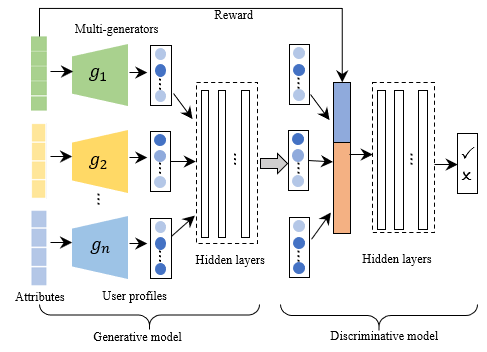}
    
  \centering{ 
      \begin{bfseries}   
          Fig. 12.
      \end{bfseries}
      LARA framework \cite{DBLP:conf/wsdm/SunLLRGN20}.}
\end{figure}

Experiments on two datasets show the effectiveness of LARA through comparisons against other state-of-the-art baselines. The same idea can be used to solve the user cold-start problem. Despite its successes, the method can still be optimized by drawing on the concept of WGAN.

[\textbf{TagRec}] Quintanilla et al. \cite{DBLP:journals/corr/abs-2004-00698} proposed an end-to-end adversarial learning framework called TagRec, to make tag recommendation. TagRec features a generator $G$ that takes image embeddings as an input and generates image tags; a discriminator $D$ is used to distinguish the generated tags from the real ones. TagRec achieved an excellent performance on two large-scale datasets. The image content and user history records involved in TagRec are applied in parallel. If the authors had attempted to design multiple generators and put both sets of information into the generator, the performance would have been enhanced.

[\textbf{RRGAN}] Similarly, Chen et al. \cite{DBLP:conf/ijcnn/ChenZWW019} proposed the rating and review GAN (RRGAN), to generate user profiles and item representations. RRGAN takes the representations learned from reviews as the input of $G$. $G$ predicts the item ratings for a given user. In contrast, $D$ attempts to distinguish the predicted representations from real ones. Furthermore, the authors plan to unify explicit and implicit feedback, using adversarial learning.

\subsection{Quantitative Analysis}
We provide an overview of the aforementioned models in Table 6, to summarize the evaluation metrics, experimental data, and dataset domains involved in the studies. Precision is the most widely adopted metric, followed by NDCG, with a partial overlap with the GAN-based models for mitigating the data sparsity issue. In terms of the dataset domains, movies are the most-used domain, owing to the popularity of Movielens. The following section presents a performance analysis of three prominent models.

\begin{table*}[pos=htp]
\small
\centering{ 
      \begin{bfseries}   
         Table 6
      \end{bfseries}
     Evaluation and domain comparison of GAN-based models for mitigating data sparsity issue. (ML: Movielens, PT: Pinterest, GW: Gowalla, AM: Amazon, NF: Netflix, FS: Foursquare, LS: Last.fm, BS: Business, IM: Image, MV: Movie, MS: Music, SO: Social, PP: Paper, BK: Book) \\}

\begin{tabular}{c|l|cccccccc|c|c}
\hline
 &
  \multicolumn{1}{c|}{} &
  \multicolumn{8}{c|}{Evaluation Metric} &
   &
   \\ \cline{3-10}
\multirow{-2}{*}{Category} &
  \multicolumn{1}{c|}{\multirow{-2}{*}{Model}} &
  \rotatebox{90}{NDCG} &
  \rotatebox{90}{HR} &
  \rotatebox{90}{Recall} &
  \rotatebox{90}{MAE} &
  \rotatebox{90}{RMSE} &
  \multicolumn{1}{l}{\rotatebox{90}{Pre}} &
  \multicolumn{1}{l}{\rotatebox{90}{MAP}} &
  \multicolumn{1}{l|}{\rotatebox{90}{AUC}} &
  \multirow{-2}{*}{Domain} &
  \multirow{-2}{*}{Datasets} \\ \hline
 &
  CFGAN \cite{DBLP:conf/cikm/ChaeKKL18} &
   &
   &
  \checkmark &
   &
   &
  \checkmark &
  \checkmark &
   &
  MV &
  Ciao, WT, ML \\
 &
  \cellcolor[HTML]{EFEFEF}RAGAN \cite{DBLP:conf/www/ChaeKKC19} &
  \cellcolor[HTML]{EFEFEF} &
  \cellcolor[HTML]{EFEFEF} &
  \cellcolor[HTML]{EFEFEF} &
  \cellcolor[HTML]{EFEFEF} &
  \cellcolor[HTML]{EFEFEF} &
  \cellcolor[HTML]{EFEFEF} \checkmark &
  \cellcolor[HTML]{EFEFEF} &
  \cellcolor[HTML]{EFEFEF} &
  \cellcolor[HTML]{EFEFEF}MV &
  \cellcolor[HTML]{EFEFEF}Ciao, WT, ML \\
 &
  AugCF \cite{DBLP:conf/kdd/WangYWNHC19} &
   &
  \checkmark &
   &
   &
   &
   &
   &
   &
  MV &
  ML, FP \\
 &
  \cellcolor[HTML]{EFEFEF}APL \cite{DBLP:journals/eswa/SunWWY19} &
  \cellcolor[HTML]{EFEFEF} \checkmark &
  \cellcolor[HTML]{EFEFEF} &
  \cellcolor[HTML]{EFEFEF} \checkmark &
  \cellcolor[HTML]{EFEFEF} &
  \cellcolor[HTML]{EFEFEF} &
  \cellcolor[HTML]{EFEFEF} \checkmark &
  \cellcolor[HTML]{EFEFEF} \checkmark &
  \cellcolor[HTML]{EFEFEF} &
  \cellcolor[HTML]{EFEFEF}\begin{tabular}[c]{@{}c@{}}IM, BS,\\ MS\end{tabular} &
  \cellcolor[HTML]{EFEFEF}\begin{tabular}[c]{@{}c@{}}GW, Yelp,\\ PT,YH\end{tabular} \\
 &
  PLASTIC \cite{DBLP:conf/ijcai/ZhaoWYGYC18} &
  \checkmark &
   &
   &
   &
   &
  \checkmark &
   &
   &
  MV &
  ML, NF \\
 &
  \cellcolor[HTML]{EFEFEF}RecGAN \cite{DBLP:conf/recsys/BharadhwajPL18} &
  \cellcolor[HTML]{EFEFEF} \checkmark &
  \cellcolor[HTML]{EFEFEF} &
  \cellcolor[HTML]{EFEFEF} &
  \cellcolor[HTML]{EFEFEF} &
  \cellcolor[HTML]{EFEFEF} &
  \cellcolor[HTML]{EFEFEF} &
  \cellcolor[HTML]{EFEFEF} \checkmark &
  \cellcolor[HTML]{EFEFEF} &
  \cellcolor[HTML]{EFEFEF}MV &
  \cellcolor[HTML]{EFEFEF}MFP, NF \\
 &
  APOIR \cite{DBLP:conf/www/0002YZTZW19} &
  \checkmark &
   &
  \checkmark &
   &
   &
  \checkmark &
  \checkmark &
   &
  POI &
  \begin{tabular}[c]{@{}c@{}}GW, Yelp,\\ FS\end{tabular} \\
 &
  \cellcolor[HTML]{EFEFEF}{\color[HTML]{333333} Geo-ALM \cite{DBLP:conf/ijcai/0061W0Y19}} &
  \cellcolor[HTML]{EFEFEF} &
  \cellcolor[HTML]{EFEFEF} &
  \cellcolor[HTML]{EFEFEF} \checkmark &
  \cellcolor[HTML]{EFEFEF} &
  \cellcolor[HTML]{EFEFEF} &
  \cellcolor[HTML]{EFEFEF} \checkmark &
  \cellcolor[HTML]{EFEFEF} &
  \cellcolor[HTML]{EFEFEF} &
  \cellcolor[HTML]{EFEFEF}POI &
  \cellcolor[HTML]{EFEFEF}GW, FS \\
\multirow{-9}{*}{\begin{tabular}[c]{@{}c@{}}Models\\ for\\ generating\\ user\\ preferences\\ by\\ augmenting\\ with\\ interactive\\ information\end{tabular}} &
  UGAN \cite{DBLP:conf/pakdd/WangGWYWX19} &
   &
   &
  \checkmark &
  \checkmark &
  \checkmark &
  \checkmark &
  \checkmark &
   &
  MV &
  ML, DB \\ \hline
 &
  \cellcolor[HTML]{EFEFEF}RSGAN \cite{DBLP:journals/corr/abs-1909-03529} &
  \cellcolor[HTML]{EFEFEF} \checkmark &
  \cellcolor[HTML]{EFEFEF} &
  \cellcolor[HTML]{EFEFEF} \checkmark &
  \cellcolor[HTML]{EFEFEF} &
  \cellcolor[HTML]{EFEFEF} &
  \cellcolor[HTML]{EFEFEF} \checkmark &
  \cellcolor[HTML]{EFEFEF} &
  \cellcolor[HTML]{EFEFEF} &
  \cellcolor[HTML]{EFEFEF}SO &
  \cellcolor[HTML]{EFEFEF}LF, DB, EP \\
 &
  KTGAN \cite{DBLP:conf/icdm/YangGWJXW18} &
  \checkmark &
   &
   &
   &
   &
  \checkmark &
   &
   &
  MV &
  ML, DB \\
 &
  \cellcolor[HTML]{EFEFEF}CnGAN \cite{DBLP:conf/www/PereraZ19} &
  \cellcolor[HTML]{EFEFEF} \checkmark &
  \cellcolor[HTML]{EFEFEF} \checkmark &
  \cellcolor[HTML]{EFEFEF} &
  \cellcolor[HTML]{EFEFEF} &
  \cellcolor[HTML]{EFEFEF} &
  \cellcolor[HTML]{EFEFEF} &
  \cellcolor[HTML]{EFEFEF} &
  \cellcolor[HTML]{EFEFEF} &
  \cellcolor[HTML]{EFEFEF}SO, MS &
  \cellcolor[HTML]{EFEFEF}YT, TW \\
 &
  RecSys-DAN \cite{DBLP:journals/corr/abs-1903-10794} &
   &
   &
   &
  \checkmark &
   &
   &
   &
   &
  MS &
  AM \\
 &
  \cellcolor[HTML]{EFEFEF}ATR \cite{DBLP:conf/sigir/RafailidisC19} &
  \cellcolor[HTML]{EFEFEF} &
  \cellcolor[HTML]{EFEFEF} &
  \cellcolor[HTML]{EFEFEF} &
  \cellcolor[HTML]{EFEFEF} &
  \cellcolor[HTML]{EFEFEF} \checkmark &
  \cellcolor[HTML]{EFEFEF} &
  \cellcolor[HTML]{EFEFEF} &
  \cellcolor[HTML]{EFEFEF} &
  \cellcolor[HTML]{EFEFEF}RW &
  \cellcolor[HTML]{EFEFEF}AM \\
 &
  DVBPR \cite{DBLP:conf/icdm/KangFWM17} &
   &
   &
   &
   &
   &
   &
   &
  \checkmark &
  RW, IM &
  AM \\
 &
  \cellcolor[HTML]{EFEFEF}LARA \cite{DBLP:conf/wsdm/SunLLRGN20} &
  \cellcolor[HTML]{EFEFEF} \checkmark &
  \cellcolor[HTML]{EFEFEF} &
  \cellcolor[HTML]{EFEFEF} &
  \cellcolor[HTML]{EFEFEF} \checkmark &
  \cellcolor[HTML]{EFEFEF} &
  \cellcolor[HTML]{EFEFEF} \checkmark &
  \cellcolor[HTML]{EFEFEF} &
  \cellcolor[HTML]{EFEFEF} &
  \cellcolor[HTML]{EFEFEF}MV, BS &
  \cellcolor[HTML]{EFEFEF}ML, Inzone \\
 &
  TagRec \cite{DBLP:journals/corr/abs-2004-00698} &
   &
   &
  \checkmark &
  \checkmark &
   &
  \checkmark &
   &
   &
  IM &
  \begin{tabular}[c]{@{}c@{}}Nus-WIDE,\\ YFCC\end{tabular} \\
 &
  \cellcolor[HTML]{EFEFEF}LSIC \cite{DBLP:conf/ijcai/FanD0WTL19} &
  \cellcolor[HTML]{EFEFEF} \checkmark &
  \cellcolor[HTML]{EFEFEF} &
  \cellcolor[HTML]{EFEFEF} &
  \cellcolor[HTML]{EFEFEF} &
  \cellcolor[HTML]{EFEFEF} &
  \cellcolor[HTML]{EFEFEF} \checkmark &
  \cellcolor[HTML]{EFEFEF} \checkmark &
  \cellcolor[HTML]{EFEFEF} &
  \cellcolor[HTML]{EFEFEF}MV &
  \cellcolor[HTML]{EFEFEF}ML, NF \\
\multirow{-10}{*}{\begin{tabular}[c]{@{}c@{}}Models\\ for \\ synthesizing\\ user\\ preferences\\ by\\ augmenting\\ with\\ auxiliary\\ information\end{tabular}} &
  RRGAN \cite{DBLP:conf/ijcnn/ChenZWW019} &
  \checkmark &
   &
  \checkmark &
   &
   &
   &
   &
   &
  BS &
  \begin{tabular}[c]{@{}c@{}}Watches, patio,\\ Kindle\end{tabular} \\ \hline
\end{tabular}
\end{table*}

We only collected the experimental results from three studies when making our comparison, because results are displayed in different forms across different studies. In the interest of fairness and accuracy, we summarized the data sparsity and dataset information by employing a similar sparsity in our comparison (Ciao and Yahoo). We found that CFGAN outperformed IRGAN and GraphGAN on the Movielens and Ciao datasets, as shown in Table 7. Furthermore, we can observe that almost all studies regard IRGAN as the basic  model. The overall performances of the three models are better than that of IRGAN.
\begin{table}[pos=htp]

\centering{ 
      \begin{bfseries}   
         Table 7
      \end{bfseries}
     Performance comparison of GAN-based models for mitigating data sparsity, for three datasets \\ (results from CFGAN \cite{DBLP:conf/cikm/ChaeKKL18}, PLASTIC \cite{DBLP:conf/ijcai/ZhaoWYGYC18}, and APL \cite{DBLP:journals/eswa/SunWWY19}).}
     
\begin{tabular}{c|c|c|ccllcl}
\hline
\multicolumn{1}{l|}{} &
   &
   &
  \multicolumn{6}{c|}{Evaluation Metric} \\ \cline{4-9} 
\multicolumn{1}{l|}{\multirow{-2}{*}{Dataset}} &
  \multirow{-2}{*}{Source} &
  \multirow{-2}{*}{Model} &
  \multicolumn{1}{l}{Pre@5} &
  \multicolumn{1}{l}{Pre@10} &
  Pre@20 &
  NDCG@5 &
  \multicolumn{1}{l}{NDCG@10} &
  NDCG@20 \\ \hline
 &
   &
  IRGAN &
  0.312 &
  / &
  0.221 &
  0.342 &
  / &
  0.368 \\
 &
   &
  \cellcolor[HTML]{EFEFEF}GraphGAN &
  \cellcolor[HTML]{EFEFEF}0.212 &
  \cellcolor[HTML]{EFEFEF}/ &
  \cellcolor[HTML]{EFEFEF}0.151 &
  \cellcolor[HTML]{EFEFEF}0.183 &
  \cellcolor[HTML]{EFEFEF}/ &
  \cellcolor[HTML]{EFEFEF}0.183 \\
 &
  \multirow{-3}{*}{\begin{tabular}[c]{@{}c@{}}CFGAN\\ \cite{DBLP:conf/cikm/ChaeKKL18} \end{tabular}} &
  CFGAN &
  0.444 &
  / &
  0.294 &
  0.476 &
  / &
  0.433 \\ \cline{2-9} 
 &
   &
  \cellcolor[HTML]{EFEFEF}IRGAN &
  \cellcolor[HTML]{EFEFEF}0.2885 &
  \cellcolor[HTML]{EFEFEF}/ &
  \multicolumn{1}{c}{\cellcolor[HTML]{EFEFEF}/} &
  \multicolumn{1}{c}{\cellcolor[HTML]{EFEFEF}0.3032} &
  \cellcolor[HTML]{EFEFEF}/ &
  \multicolumn{1}{c}{\cellcolor[HTML]{EFEFEF}/} \\
\multirow{-5}{*}{\begin{tabular}[c]{@{}c@{}}Movielens\\ -100K\\ (93.69\%)\end{tabular}} &
  \multirow{-2}{*}{\begin{tabular}[c]{@{}c@{}}PLASTIC\\ \cite{DBLP:conf/ijcai/ZhaoWYGYC18} \end{tabular}} &
  PLASTIC &
  0.3115 &
  / &
  \multicolumn{1}{c}{/} &
  \multicolumn{1}{c}{0.3312} &
  / &
  \multicolumn{1}{c}{/} \\ \hline
 &
   &
  \cellcolor[HTML]{EFEFEF}IRGAN &
  \cellcolor[HTML]{EFEFEF}0.035 &
  \cellcolor[HTML]{EFEFEF}/ &
  \cellcolor[HTML]{EFEFEF}0.023 &
  \cellcolor[HTML]{EFEFEF}0.046 &
  \cellcolor[HTML]{EFEFEF}/ &
  \cellcolor[HTML]{EFEFEF}0.066 \\
 &
   &
  GraphGAN &
  0.026 &
  / &
  0.017 &
  0.041 &
  / &
  0.058 \\
\multirow{-3}{*}{\begin{tabular}[c]{@{}c@{}}Ciao\\ (98.72\%)\end{tabular}} &
  \multirow{-3}{*}{\begin{tabular}[c]{@{}c@{}}CFGAN\\ \cite{DBLP:conf/cikm/ChaeKKL18} \end{tabular}} &
  \cellcolor[HTML]{EFEFEF}CFGAN &
  \cellcolor[HTML]{EFEFEF}0.072 &
  \cellcolor[HTML]{EFEFEF}/ &
  \cellcolor[HTML]{EFEFEF}0.045 &
  \cellcolor[HTML]{EFEFEF}0.092 &
  \cellcolor[HTML]{EFEFEF}/ &
  \cellcolor[HTML]{EFEFEF}0.124 \\ \hline
 &
   &
  IRGAN &
  / &
  10.960 &
  \multicolumn{1}{c}{/} &
  \multicolumn{1}{c}{/} &
  11.900 &
  \multicolumn{1}{c}{/} \\
\multirow{-2}{*}{\begin{tabular}[c]{@{}c@{}}Yahoo\\ (99.22\%)\end{tabular}} &
  \multirow{-2}{*}{\begin{tabular}[c]{@{}c@{}}APL\\ \cite{DBLP:journals/eswa/SunWWY19} \end{tabular}} &
  \cellcolor[HTML]{EFEFEF}APL &
  \cellcolor[HTML]{EFEFEF}/ &
  \cellcolor[HTML]{EFEFEF}11.513 &
  \multicolumn{1}{c}{\cellcolor[HTML]{EFEFEF}/} &
  \multicolumn{1}{c}{\cellcolor[HTML]{EFEFEF}/} &
  \cellcolor[HTML]{EFEFEF}12.661 &
  \multicolumn{1}{c}{\cellcolor[HTML]{EFEFEF}/} \\ \hline
\end{tabular}
\end{table}

\subsection{Qualitative Analysis}
To better highlight the differences between the aforementioned models in term of their generator and discriminators, we list their specific designs and respective advantages in Table 8.

\begin{table*}[pos=htp]

\centering{ 
      \begin{bfseries}   
         Table 8
      \end{bfseries}
     Schematic representation of GAN-based RS designed to mitigate the data sparsity issue.}
     
\begin{tabular}{|c|c|l|l|l|}
\hline
Category &
  Model &
  \multicolumn{1}{c|}{Generator} &
  \multicolumn{1}{c|}{Discriminator} &
  \multicolumn{1}{c|}{Advantage} \\ \hline
\multirow{8}{*}{\begin{tabular}[c]{@{}c@{}}Models\\ for\\ generating \\ user\\ preferences \\ by \\ augmenting\\ with\\ interactive\\ information\end{tabular}} &
  \begin{tabular}[c]{@{}c@{}}CFGAN\\ \cite{DBLP:conf/cikm/ChaeKKL18} \end{tabular} &
  \begin{tabular}[c]{@{}l@{}}Apply user purchase\\ vectors to generate \\ purchase vectors\end{tabular} &
  \begin{tabular}[c]{@{}l@{}}Identify whether the\\ purchase vectors meet\\ the users' preferences\end{tabular} &
  \begin{tabular}[c]{@{}l@{}}Using vector-wise\\ training in RS\end{tabular} \\ \cline{2-5} 
 &
  \begin{tabular}[c]{@{}c@{}}AugCF\\ \cite{DBLP:conf/kdd/WangYWNHC19} \end{tabular} &
  \begin{tabular}[c]{@{}l@{}}Apply interactive \\ categories as \\ condition label to \\ generate items\end{tabular} &
  \begin{tabular}[c]{@{}l@{}}Two stages: (1) \\ determine the relationships \\ between users and items,\\ and (2) identify \\ the relationship labels\end{tabular} &
  \begin{tabular}[c]{@{}l@{}}Design the \\ discriminator to\\ identify true/false\\ data and like/dislike\\ categories\end{tabular} \\ \cline{2-5} 
 &
  \begin{tabular}[c]{@{}c@{}}APL\\ \cite{DBLP:journals/eswa/SunWWY19} \end{tabular} &
  \begin{tabular}[c]{@{}l@{}}Generate the items\\ to approximate the\\ actual distribution\end{tabular} &
  \begin{tabular}[c]{@{}l@{}}Apply the pair-wise \\ ranking function to \\ determine\end{tabular} &
  \begin{tabular}[c]{@{}l@{}}Apply adversarial \\ learning to the\\ implicit feedback\end{tabular} \\ \cline{2-5} 
 &
  \begin{tabular}[c]{@{}c@{}}PLASTIC\\ \cite{DBLP:conf/ijcai/ZhaoWYGYC18} \end{tabular} &
  \begin{tabular}[c]{@{}l@{}}Apply time to \\ generate a list of \\ items\end{tabular} &
  \begin{tabular}[c]{@{}l@{}}Capture long-term and \\ short-term preferences of\\ users to select the exact\\ high-scoring items\end{tabular} &
  \begin{tabular}[c]{@{}l@{}}Apply adversarial\\ learning to combine \\ MF and RNN\end{tabular} \\ \cline{2-5} 
 &
  \begin{tabular}[c]{@{}c@{}}APOIR\\\cite{DBLP:conf/www/0002YZTZW19} \end{tabular} &
  \begin{tabular}[c]{@{}l@{}}Select POIs to fit \\ the real data \\ distribution\end{tabular} &
  \begin{tabular}[c]{@{}l@{}}Determine whether user's\\ POIs are real or \\ generated\end{tabular} &
  \begin{tabular}[c]{@{}l@{}}Apply adversarial\\ learning to optimize \\ POI RS\end{tabular} \\ \cline{2-5} 
 &
  \begin{tabular}[c]{@{}c@{}}RecGAN\\ \cite{DBLP:conf/recsys/BharadhwajPL18} \end{tabular} &
  \begin{tabular}[c]{@{}l@{}}Predict a sequence \\ of items consumed\\ by user\end{tabular} &
  \begin{tabular}[c]{@{}l@{}}A GRU that determines the\\ probability of items\\ sequences\end{tabular} &
  \begin{tabular}[c]{@{}l@{}}Combine GRU and GANs \\ to capture temporal \\ profiles\end{tabular} \\ \cline{2-5} 
 &
  \begin{tabular}[c]{@{}c@{}}Geo-ALM\\ \cite{DBLP:conf/ijcai/0061W0Y19} \end{tabular} &
  \begin{tabular}[c]{@{}l@{}}Generate high-quality\\ unvisited POIs for a\\ given user\end{tabular} &
  \begin{tabular}[c]{@{}l@{}}Distinguish between \\ visited POIs and \\ unvisited POIs\end{tabular} &
  \begin{tabular}[c]{@{}l@{}}Combine geographic\\ features and GANs\end{tabular} \\ \cline{2-5} 
 &
  \begin{tabular}[c]{@{}c@{}}UGAN\\ \cite{DBLP:conf/pakdd/WangGWYWX19} \end{tabular} &
  \begin{tabular}[c]{@{}l@{}}Simulate user rating\\ information\end{tabular} &
  \begin{tabular}[c]{@{}l@{}}Distinguish between \\ generated and real data\end{tabular} &
  \begin{tabular}[c]{@{}l@{}}Use wGAN and CGAN to\\ forge users\end{tabular} \\ \hline
\multirow{8}{*}{\begin{tabular}[c]{@{}c@{}}Models\\ for\\ synthesizing\\ user\\ preferences\\ by\\ augmenting\\ with\\ auxiliary\\ information\end{tabular}} &
  \begin{tabular}[c]{@{}c@{}}RSGAN\\ \cite{DBLP:journals/corr/abs-1909-03529} \end{tabular} &
  \begin{tabular}[c]{@{}l@{}}Generate reliable\\ friends and their\\ consumed items using\\ Gumbel-Softmax\end{tabular} &
  \begin{tabular}[c]{@{}l@{}}Apply the positive,\\ negative, and generated \\ items to sort \\ the candidate items\end{tabular} &
  \begin{tabular}[c]{@{}l@{}}Apply GANs to \\ social recommendation\end{tabular} \\ \cline{2-5} 
 &
  \begin{tabular}[c]{@{}c@{}}KTGAN\\ \cite{DBLP:conf/icdm/YangGWJXW18} \end{tabular} &
  \begin{tabular}[c]{@{}l@{}}Apply the auxiliary \\ information to generate\\ relationships between \\ users and items\end{tabular} &
  \begin{tabular}[c]{@{}l@{}}Distinguish whether the \\ relationship pair \\ is from a real dataset\end{tabular} &
  \begin{tabular}[c]{@{}l@{}}Integrate auxiliary \\ information and\\ IRGAN to alleviate \\ data sparsity\end{tabular} \\ \cline{2-5} 
 &
  \begin{tabular}[c]{@{}c@{}}CnGAN\\ \cite{DBLP:conf/www/PereraZ19} \end{tabular} &
  \begin{tabular}[c]{@{}l@{}}Generate the mapping\\ encoding of the source\\ network for the \\ non-overlapping users\end{tabular} &
  \begin{tabular}[c]{@{}l@{}}Determine whether the mapping \\ relationship of\\ the overlapping users is\\ real or generated\end{tabular} &
  \begin{tabular}[c]{@{}l@{}}Apply GANs to \\ the cross-domain RS\end{tabular} \\ \cline{2-5} 
 &
  \begin{tabular}[c]{@{}c@{}}RecSys-DAN\\ \cite{DBLP:journals/corr/abs-1903-10794} \end{tabular} &
  \begin{tabular}[c]{@{}l@{}}Learn the vectors\\ from different domains \\ to learn the transfer \\ mapping\end{tabular} &
  \begin{tabular}[c]{@{}l@{}}Determine whether the relationships\\ between users and items \\ in different fields are \\ genuine or not\end{tabular} &
  \begin{tabular}[c]{@{}l@{}}Learn how to represent\\ users, items, and their \\ interactions in\\ different domains\end{tabular} \\ \cline{2-5} 
 &
  \begin{tabular}[c]{@{}c@{}}ATR\\ \cite{DBLP:conf/sigir/RafailidisC19} \end{tabular} &
  \begin{tabular}[c]{@{}l@{}}Use the encoder-decoder\\ to generate reviews\end{tabular} &
  \begin{tabular}[c]{@{}l@{}}Estimate the probability\\ that the review is true\end{tabular} &
  \begin{tabular}[c]{@{}l@{}}An adversarial model\\ for review-based \\ recommendations\end{tabular} \\ \cline{2-5} 
 &
  \begin{tabular}[c]{@{}c@{}}LARA\\ \cite{DBLP:conf/wsdm/SunLLRGN20} \end{tabular} &
  \begin{tabular}[c]{@{}l@{}}Generate the profiles of\\ the user from different\\ attributes\end{tabular} &
  \begin{tabular}[c]{@{}l@{}}Distinguish the well-\\ matched user-item pairs\\ from ill-matched ones\end{tabular} &
  \begin{tabular}[c]{@{}l@{}}A GAN-based model that\\ uses multi-generators\end{tabular} \\ \cline{2-5} 
 &
  \begin{tabular}[c]{@{}c@{}}TagRec\\ \cite{DBLP:journals/corr/abs-2004-00698} \end{tabular} &
  \begin{tabular}[c]{@{}l@{}}Generate (predict) \\ personalized tags\end{tabular} &
  \begin{tabular}[c]{@{}l@{}}Distinguish the generated\\ tags from real ones\end{tabular} &
  \begin{tabular}[c]{@{}l@{}}Use GANs to predict\\ tags resembling truth\\ tags\end{tabular} \\ \cline{2-5} 
 &
  \begin{tabular}[c]{@{}c@{}}RRGAN\\ \cite{DBLP:conf/ijcnn/ChenZWW019} \end{tabular} &
  \begin{tabular}[c]{@{}l@{}}Generate vectors of\\ users and items based\\ on reviews\end{tabular} &
  \begin{tabular}[c]{@{}l@{}}Distinguish the predicted\\ ratings from real ones\end{tabular} &
  \begin{tabular}[c]{@{}l@{}}Predict the ratings \\ based on reviews\end{tabular} \\ \hline
\end{tabular}
\end{table*}

\section{Open Issues and Future Research Directions}
\subsection{Open Issues}
\begin{enumerate}[\textbullet]
\item \textbf{The Position of Adversarial Training.}
DNN-based recommendation models have recently been extensively researched, because they can learn more abstract representations of users and items and can grasp the nonlinear structural features of interactive information. However, these networks feature complex structures and a wide variety of parameters. Hence, choosing a suitable adversarial training position has become a significant challenge, requiring more in-depth knowledge and broader exploration.

\item \textbf{Model Parameter Optimization Stability Problem for Discrete Training Data.}
GANs were initially designed for the image domain, where data are  continuous and gradients are applied for differentiable values. However, the interactive records in RSs are discrete. The generation of recommendation lists is a sampling operation. Thus, the gradients derived from the objective functions of the original GANs cannot be directly used to optimize the generator via gradient descent. This problem prevents the model parameters from converging during training. Several researchers have tried to train the model, using a policy gradient \cite{DBLP:journals/ml/Williams92}, Gumbel--Softmax \cite{DBLP:conf/iclr/JangGP17}, and Rao--Blackwell sampling \cite{DBLP:conf/flairs/CostaD19}; however, the stability of the parameter optimization in the GAN-based recommendation model is still an open research problem.

\item \textbf{Model Size and Time Complexity.}
The scale and time complexities of the existing GAN-based RSs are relatively high, because GANs include two components, with each part consisting of multiple layers of neural networks. This problem becomes more serious when the model contains multiple generators and discriminators, and it hinders the application of GAN-based recommendation models to real systems. Besides this, the training process is confrontational and mutually promoting, which can be considered as a minimax game. We must ensure high-quality feedback between the two modules to enhance the model performance. If one of the two models is under-fitting, it will cause the other  model to collapse.
\end{enumerate}

\subsection{Future Research Directions}
Although studies on GAN-based recommendation models have established a solid foundation for alleviating the data noise and data sparsity issues, several open issues remain. In this section, we put forward the more promising areas of interest in GAN-based RS research and introduce the following future research directions:

\begin{enumerate}[\textbullet]
\item \textbf{GAN-Based Explanations for Recommendation Models.}
The generators and discriminators in GAN-based recommendation models are predominately constructed using DNNs, which are categorized as black-box models. In other words, we are only aware of their inputs and outputs, and the underlying principle is difficult to understand. Existing models that improve the interpretability of RSs primarily give explanations following a recommendation \cite{DBLP:conf/aaai/WangWX00C19, DBLP:conf/wsdm/SinghA19}, and the content of the explanation is often unrelated to the results. A beneficial development would be to use GANs to explain the results and generate a recommendation list. In this framework, the discriminator not only judges the accuracy of the generated recommendation but also the interpretation of this recommendation. Thus, the generator and discriminator can compete with each other, to improve model explainability.

\item \textbf{Cross-Domain Recommendation Based on GANs.}
Cross-domain models, which assist in the representation of the target domain using the knowledge learned from source domains, are a promising option for tackling the data sparsity issue. One of the most widely studied topics in cross-domain recommendation is transfer learning \cite{DBLP:journals/corr/abs-1901-06125}. This aims to improve the learning ability in one target domain by using the knowledge transferred from other domains. However, unifying the information from different domains into the same representation space remains a challenging problem. Adversarial learning can continuously learn and optimize the mapping process from the source domain to the target one, thereby enriching the training data of the recommendation model. A small number of models \cite{DBLP:conf/www/PereraZ19, DBLP:journals/corr/abs-1903-10794} have employed the advantages of GANs in learning the mapping relationship between different fields of information, and their effectiveness has been verified through experiments. This is a promising but mostly under-explored area, where more studies are expected.

\item \textbf{Scalability of GAN-Based Recommendation Models.}
Scalability is critical for recommendation models, because the ever-increasing volumes of data make the time complexity a principal consideration. GANs have been applied to some commercial products; however, due to the continuous improvement of GPU computing power, further research on GAN-based RSs is required in three areas: (1) incremental learning of non-stationary and streaming data, such as the case in which large numbers of interactions take place between users and items; (2) accurate calculation of high-dimensional tensors and multimedia data sources; (3) balancing model complexity and scalability under the exponential growth of parameters. Knowledge distillation \cite{DBLP:journals/tois/ChenZXQZ19, DBLP:journals/tmm/DuFYXCT19} is an ideal candidate method to manage these areas; it utilizes a small student model that receives knowledge from a teacher model. Because short training times are essential for real-time applications, scalability is another promising direction that deserves further study.
\end{enumerate}

\section{Conclusion}
In this paper, we provided a retrospective review of the up-to-date GAN-based recommendation models, demonstrating their ability to reduce the adverse effects of data noise and alleviate the data sparsity problem. We discussed the development history of GANs and clarified their feasibility for RSs. In terms of the efforts devoted to tackling data noise, we introduced existing models from two perspectives: (1) models for mitigating malicious noise; and (2) models for distinguishing informative samples from unobserved items. In terms of the studies focusing on mitigating the data sparsity problem, we grouped the models into two categories: (1) models for generating user preferences through augmentation with interactive information; and (2) models for synthesizing user preferences through augmentation with auxiliary information. After the review, we discussed some of the most significant open problems and highlighted several promising future directions. We hope this survey will help shape the ideas of researchers and provide some practical guidelines for this new discipline.

\section{Acknowledgements}
This research was supported by the National Key Research and Development Program of China (2018YFF0214706), the Graduate Scientific Research and Innovation Foundation of Chongqing, China (CYS19028), the Natural Science Foundation of Chongqing, China (cstc2020jcyj-msxmX06), and the Fundamental Research Funds for the Central Universities of Chongqing University  (2020CDJ-LHZZ-039).

\bibliographystyle{cas-model2-names}

\bibliography{cas-survey}

\begin{thebibliography}{84}
\expandafter\ifx\csname natexlab\endcsname\relax\def\natexlab#1{#1}\fi
\providecommand{\url}[1]{\texttt{#1}}
\providecommand{\href}[2]{#2}
\providecommand{\path}[1]{#1}
\providecommand{\DOIprefix}{doi:}
\providecommand{\ArXivprefix}{arXiv:}
\providecommand{\URLprefix}{URL: }
\providecommand{\Pubmedprefix}{pmid:}
\providecommand{\doi}[1]{\href{http://dx.doi.org/#1}{\path{#1}}}
\providecommand{\Pubmed}[1]{\href{pmid:#1}{\path{#1}}}
\providecommand{\bibinfo}[2]{#2}
\ifx\xfnm\relax \def\xfnm[#1]{\unskip,\space#1}\fi
\bibitem[{Arjovsky and Bottou(2017)}]{DBLP:conf/iclr/ArjovskyB17}
\bibinfo{author}{Arjovsky, M.}, \bibinfo{author}{Bottou, L.},
  \bibinfo{year}{2017}.
\newblock \bibinfo{title}{Towards principled methods for training generative
  adversarial networks}, in: \bibinfo{booktitle}{5th International Conference
  on Learning Representations, {ICLR} 2017, Toulon, France, April 24-26, 2017},
  pp. \bibinfo{pages}{57--57}.
\bibitem[{Arjovsky et~al.(2017)Arjovsky, Chintala and
  Bottou}]{DBLP:conf/icml/ArjovskyCB17}
\bibinfo{author}{Arjovsky, M.}, \bibinfo{author}{Chintala, S.},
  \bibinfo{author}{Bottou, L.}, \bibinfo{year}{2017}.
\newblock \bibinfo{title}{Wasserstein generative adversarial networks}, in:
  \bibinfo{booktitle}{Proceedings of the 34th International Conference on
  Machine Learning, {ICML} 2017, Sydney, NSW, Australia, August 6-11, 2017},
  pp. \bibinfo{pages}{214--223}.
\bibitem[{Bag et~al.(2019)Bag, Kumar, Awasthi and
  Tiwari}]{DBLP:journals/dss/BagKAT19}
\bibinfo{author}{Bag, S.}, \bibinfo{author}{Kumar, S.},
  \bibinfo{author}{Awasthi, A.}, \bibinfo{author}{Tiwari, M.K.},
  \bibinfo{year}{2019}.
\newblock \bibinfo{title}{A noise correction-based approach to support a
  recommender system in a highly sparse rating environment}.
\newblock \bibinfo{journal}{Decision Support Systems} \bibinfo{volume}{118},
  \bibinfo{pages}{46--57}.
\bibitem[{Baldwin and Chai(2012)}]{DBLP:conf/naacl/BaldwinC12}
\bibinfo{author}{Baldwin, T.}, \bibinfo{author}{Chai, J.Y.},
  \bibinfo{year}{2012}.
\newblock \bibinfo{title}{Autonomous self-assessment of autocorrections:
  Exploring text message dialogues}, in: \bibinfo{booktitle}{Human Language
  Technologies: Conference of the North American Chapter of the Association of
  Computational Linguistics, Montr{\'{e}}al, Canada, June 3-8, 2012}, pp.
  \bibinfo{pages}{710--719}.
\bibitem[{Beigi et~al.(2020)Beigi, Mosallanezhad, Guo, Alvari, Nou and
  Liu}]{DBLP:conf/wsdm/BeigiMGAN020}
\bibinfo{author}{Beigi, G.}, \bibinfo{author}{Mosallanezhad, A.},
  \bibinfo{author}{Guo, R.}, \bibinfo{author}{Alvari, H.},
  \bibinfo{author}{Nou, A.}, \bibinfo{author}{Liu, H.}, \bibinfo{year}{2020}.
\newblock \bibinfo{title}{Privacy-aware recommendation with private-attribute
  protection using adversarial learning}, in: \bibinfo{booktitle}{{WSDM} '20:
  The Thirteenth {ACM} International Conference on Web Search and Data Mining,
  Houston, TX, USA, February 3-7, 2020}, pp. \bibinfo{pages}{34--42}.
\bibitem[{Bharadhwaj et~al.(2018)Bharadhwaj, Park and
  Lim}]{DBLP:conf/recsys/BharadhwajPL18}
\bibinfo{author}{Bharadhwaj, H.}, \bibinfo{author}{Park, H.},
  \bibinfo{author}{Lim, B.Y.}, \bibinfo{year}{2018}.
\newblock \bibinfo{title}{Recgan: recurrent generative adversarial networks for
  recommendation systems}, in: \bibinfo{booktitle}{Proceedings of the 12th
  {ACM} Conference on Recommender Systems, RecSys 2018, Vancouver, BC, Canada,
  October 2-7, 2018}, pp. \bibinfo{pages}{372--376}.
\bibitem[{Cai et~al.(2018)Cai, Han and Yang}]{DBLP:conf/aaai/CaiHY18}
\bibinfo{author}{Cai, X.}, \bibinfo{author}{Han, J.}, \bibinfo{author}{Yang,
  L.}, \bibinfo{year}{2018}.
\newblock \bibinfo{title}{Generative adversarial network based heterogeneous
  bibliographic network representation for personalized citation
  recommendation}, in: \bibinfo{booktitle}{Proceedings of the Thirty-Second
  {AAAI} Conference on Artificial Intelligence, (AAAI-18), New Orleans,
  Louisiana, USA, February 2-7, 2018}, pp. \bibinfo{pages}{5747--5754}.
\bibitem[{Chae et~al.(2019)Chae, Kang, Kim and Choi}]{DBLP:conf/www/ChaeKKC19}
\bibinfo{author}{Chae, D.}, \bibinfo{author}{Kang, J.}, \bibinfo{author}{Kim,
  S.}, \bibinfo{author}{Choi, J.}, \bibinfo{year}{2019}.
\newblock \bibinfo{title}{Rating augmentation with generative adversarial
  networks towards accurate collaborative filtering}, in:
  \bibinfo{booktitle}{The World Wide Web Conference, {WWW} 2019, San Francisco,
  CA, USA, May 13-17, 2019}, pp. \bibinfo{pages}{2616--2622}.
\bibitem[{Chae et~al.(2018)Chae, Kang, Kim and Lee}]{DBLP:conf/cikm/ChaeKKL18}
\bibinfo{author}{Chae, D.}, \bibinfo{author}{Kang, J.}, \bibinfo{author}{Kim,
  S.}, \bibinfo{author}{Lee, J.}, \bibinfo{year}{2018}.
\newblock \bibinfo{title}{{CFGAN:} {A} generic collaborative filtering
  framework based on generative adversarial networks}, in:
  \bibinfo{booktitle}{Proceedings of the 27th {ACM} International Conference on
  Information and Knowledge Management, {CIKM} 2018, Torino, Italy, October
  22-26, 2018}, pp. \bibinfo{pages}{137--146}.
\bibitem[{Chen et~al.(2019a)Chen, Ong and
  Menon}]{DBLP:journals/corr/abs-1901-06125}
\bibinfo{author}{Chen, D.}, \bibinfo{author}{Ong, C.S.},
  \bibinfo{author}{Menon, A.K.}, \bibinfo{year}{2019}a.
\newblock \bibinfo{title}{Cold-start playlist recommendation with multitask
  learning}.
\newblock \bibinfo{journal}{CoRR} \bibinfo{volume}{abs/1901.06125}.
\bibitem[{Chen and Li(2019)}]{DBLP:conf/recsys/ChenL19}
\bibinfo{author}{Chen, H.}, \bibinfo{author}{Li, J.}, \bibinfo{year}{2019}.
\newblock \bibinfo{title}{Adversarial tensor factorization for context-aware
  recommendation}, in: \bibinfo{booktitle}{Proceedings of the 13th {ACM}
  Conference on Recommender Systems, RecSys 2019, Copenhagen, Denmark,
  September 16-20, 2019}, pp. \bibinfo{pages}{363--367}.
\bibitem[{Chen et~al.(2017)Chen, Sun, Shi and Hong}]{DBLP:conf/kdd/ChenSSH17}
\bibinfo{author}{Chen, T.}, \bibinfo{author}{Sun, Y.}, \bibinfo{author}{Shi,
  Y.}, \bibinfo{author}{Hong, L.}, \bibinfo{year}{2017}.
\newblock \bibinfo{title}{On sampling strategies for neural network-based
  collaborative filtering}, in: \bibinfo{booktitle}{Proceedings of the 23rd
  {ACM} {SIGKDD} International Conference on Knowledge Discovery and Data
  Mining, Halifax, NS, Canada, August 13 - 17, 2017}, pp.
  \bibinfo{pages}{767--776}.
\bibitem[{Chen et~al.(2019b)Chen, Zheng, Wang, Wang and
  Zhang}]{DBLP:conf/ijcnn/ChenZWW019}
\bibinfo{author}{Chen, W.}, \bibinfo{author}{Zheng, H.}, \bibinfo{author}{Wang,
  Y.}, \bibinfo{author}{Wang, W.}, \bibinfo{author}{Zhang, R.},
  \bibinfo{year}{2019}b.
\newblock \bibinfo{title}{Utilizing generative adversarial networks for
  recommendation based on ratings and reviews}, in:
  \bibinfo{booktitle}{International Joint Conference on Neural Networks,
  {IJCNN} 2019 Budapest, Hungary, July 14-19, 2019}, pp. \bibinfo{pages}{1--8}.
\bibitem[{Chen et~al.(2016)Chen, Duan, Houthooft, Schulman, Sutskever and
  Abbeel}]{DBLP:conf/nips/ChenCDHSSA16}
\bibinfo{author}{Chen, X.}, \bibinfo{author}{Duan, Y.},
  \bibinfo{author}{Houthooft, R.}, \bibinfo{author}{Schulman, J.},
  \bibinfo{author}{Sutskever, I.}, \bibinfo{author}{Abbeel, P.},
  \bibinfo{year}{2016}.
\newblock \bibinfo{title}{Infogan: Interpretable representation learning by
  information maximizing generative adversarial nets}, in:
  \bibinfo{booktitle}{Advances in Neural Information Processing Systems 29:
  Annual Conference on Neural Information Processing Systems 2016, Barcelona,
  Spain, December 5-10, 2016}, pp. \bibinfo{pages}{2172--2180}.
\bibitem[{Chen et~al.(2019c)Chen, Zhang, Xu, Qin and
  Zha}]{DBLP:journals/tois/ChenZXQZ19}
\bibinfo{author}{Chen, X.}, \bibinfo{author}{Zhang, Y.}, \bibinfo{author}{Xu,
  H.}, \bibinfo{author}{Qin, Z.}, \bibinfo{author}{Zha, H.},
  \bibinfo{year}{2019}c.
\newblock \bibinfo{title}{Adversarial distillation for efficient recommendation
  with external knowledge}.
\newblock \bibinfo{journal}{{ACM} Trans. Inf. Syst.} \bibinfo{volume}{37},
  \bibinfo{pages}{12:1--12:28}.
\bibitem[{Cheng et~al.(2018)Cheng, Ding, Zhu and
  Kankanhalli}]{DBLP:conf/www/ChengDZK18}
\bibinfo{author}{Cheng, Z.}, \bibinfo{author}{Ding, Y.}, \bibinfo{author}{Zhu,
  L.}, \bibinfo{author}{Kankanhalli, M.S.}, \bibinfo{year}{2018}.
\newblock \bibinfo{title}{Aspect-aware latent factor model: Rating prediction
  with ratings and reviews}, in: \bibinfo{booktitle}{Proceedings of the 2018
  World Wide Web Conference on World Wide Web, {WWW} 2018, Lyon, France, April
  23-27, 2018}, pp. \bibinfo{pages}{639--648}.
\bibitem[{Choi et~al.(2018)Choi, Choi, Kim, Ha, Kim and
  Choo}]{DBLP:conf/cvpr/ChoiCKH0C18}
\bibinfo{author}{Choi, Y.}, \bibinfo{author}{Choi, M.}, \bibinfo{author}{Kim,
  M.}, \bibinfo{author}{Ha, J.}, \bibinfo{author}{Kim, S.},
  \bibinfo{author}{Choo, J.}, \bibinfo{year}{2018}.
\newblock \bibinfo{title}{Stargan: Unified generative adversarial networks for
  multi-domain image-to-image translation}, in: \bibinfo{booktitle}{2018 {IEEE}
  Conference on Computer Vision and Pattern Recognition, {CVPR} 2018, Salt Lake
  City, UT, USA, June 18-22, 2018}, pp. \bibinfo{pages}{8789--8797}.
\bibitem[{Costa and Dolog(2019)}]{DBLP:conf/flairs/CostaD19}
\bibinfo{author}{Costa, F.S.D.}, \bibinfo{author}{Dolog, P.},
  \bibinfo{year}{2019}.
\newblock \bibinfo{title}{Convolutional adversarial latent factor model for
  recommender system}, in: \bibinfo{booktitle}{Proceedings of the Thirty-Second
  International Florida Artificial Intelligence Research Society Conference,
  Sarasota, Florida, USA, May 19-22 2019}, pp. \bibinfo{pages}{419--424}.
\bibitem[{Da'u et~al.(2020)Da'u, Salim, Rabiu and
  Osman}]{DBLP:journals/isci/DauSRO20}
\bibinfo{author}{Da'u, A.}, \bibinfo{author}{Salim, N.},
  \bibinfo{author}{Rabiu, I.}, \bibinfo{author}{Osman, A.},
  \bibinfo{year}{2020}.
\newblock \bibinfo{title}{Recommendation system exploiting aspect-based opinion
  mining with deep learning method}.
\newblock \bibinfo{journal}{Inf. Sci.} \bibinfo{volume}{512},
  \bibinfo{pages}{1279--1292}.
\bibitem[{Denton et~al.(2015)Denton, Chintala, Szlam and
  Fergus}]{DBLP:conf/nips/DentonCSF15}
\bibinfo{author}{Denton, E.L.}, \bibinfo{author}{Chintala, S.},
  \bibinfo{author}{Szlam, A.}, \bibinfo{author}{Fergus, R.},
  \bibinfo{year}{2015}.
\newblock \bibinfo{title}{Deep generative image models using a laplacian
  pyramid of adversarial networks}, in: \bibinfo{booktitle}{Advances in Neural
  Information Processing Systems 28: Annual Conference on Neural Information
  Processing Systems 2015, Montreal, Quebec, Canada, December 7-12, 2015}, pp.
  \bibinfo{pages}{1486--1494}.
\bibitem[{Du et~al.(2019)Du, Fang, Yi, Xu, Cheng and
  Tao}]{DBLP:journals/tmm/DuFYXCT19}
\bibinfo{author}{Du, Y.}, \bibinfo{author}{Fang, M.}, \bibinfo{author}{Yi, J.},
  \bibinfo{author}{Xu, C.}, \bibinfo{author}{Cheng, J.}, \bibinfo{author}{Tao,
  D.}, \bibinfo{year}{2019}.
\newblock \bibinfo{title}{Enhancing the robustness of neural collaborative
  filtering systems under malicious attacks}.
\newblock \bibinfo{journal}{{IEEE} Trans. Multimedia} \bibinfo{volume}{21},
  \bibinfo{pages}{555--565}.
\bibitem[{Fan et~al.(2019)Fan, Derr, Ma, Wang, Tang and
  Li}]{DBLP:conf/ijcai/FanD0WTL19}
\bibinfo{author}{Fan, W.}, \bibinfo{author}{Derr, T.}, \bibinfo{author}{Ma,
  Y.}, \bibinfo{author}{Wang, J.}, \bibinfo{author}{Tang, J.},
  \bibinfo{author}{Li, Q.}, \bibinfo{year}{2019}.
\newblock \bibinfo{title}{Deep adversarial social recommendation}, in:
  \bibinfo{booktitle}{Proceedings of the Twenty-Eighth International Joint
  Conference on Artificial Intelligence, {IJCAI} 2019, Macao, China, August
  10-16, 2019}, pp. \bibinfo{pages}{1351--1357}.
\bibitem[{Goodfellow et~al.(2014)Goodfellow, Pouget{-}Abadie, Mirza, Xu,
  Warde{-}Farley, Ozair, Courville and
  Bengio}]{DBLP:conf/nips/GoodfellowPMXWOCB14}
\bibinfo{author}{Goodfellow, I.J.}, \bibinfo{author}{Pouget{-}Abadie, J.},
  \bibinfo{author}{Mirza, M.}, \bibinfo{author}{Xu, B.},
  \bibinfo{author}{Warde{-}Farley, D.}, \bibinfo{author}{Ozair, S.},
  \bibinfo{author}{Courville, A.C.}, \bibinfo{author}{Bengio, Y.},
  \bibinfo{year}{2014}.
\newblock \bibinfo{title}{Generative adversarial nets}, in:
  \bibinfo{booktitle}{Advances in Neural Information Processing Systems 27:
  Annual Conference on Neural Information Processing Systems 2014, Montreal,
  Quebec, Canada, December 8-13, 2014}, pp. \bibinfo{pages}{2672--2680}.
\bibitem[{Goodfellow et~al.(2015)Goodfellow, Shlens and
  Szegedy}]{DBLP:journals/corr/GoodfellowSS14}
\bibinfo{author}{Goodfellow, I.J.}, \bibinfo{author}{Shlens, J.},
  \bibinfo{author}{Szegedy, C.}, \bibinfo{year}{2015}.
\newblock \bibinfo{title}{Explaining and harnessing adversarial examples}, in:
  \bibinfo{booktitle}{3rd International Conference on Learning Representations,
  {ICLR} 2015, San Diego, CA, USA, May 7-9, 2015}, pp. \bibinfo{pages}{59--66}.
\bibitem[{Gulrajani et~al.(2017)Gulrajani, Ahmed, Arjovsky, Dumoulin and
  Courville}]{DBLP:conf/nips/GulrajaniAADC17}
\bibinfo{author}{Gulrajani, I.}, \bibinfo{author}{Ahmed, F.},
  \bibinfo{author}{Arjovsky, M.}, \bibinfo{author}{Dumoulin, V.},
  \bibinfo{author}{Courville, A.C.}, \bibinfo{year}{2017}.
\newblock \bibinfo{title}{Improved training of wasserstein gans}, in:
  \bibinfo{booktitle}{Advances in Neural Information Processing Systems 30:
  Annual Conference on Neural Information Processing Systems 2017, Long Beach,
  CA, {USA}, December 4-9, 2017}, pp. \bibinfo{pages}{5767--5777}.
\bibitem[{Harper and Konstan(2016)}]{DBLP:journals/tiis/HarperK16}
\bibinfo{author}{Harper, F.M.}, \bibinfo{author}{Konstan, J.A.},
  \bibinfo{year}{2016}.
\newblock \bibinfo{title}{The movielens datasets: History and context}.
\newblock \bibinfo{journal}{TiiS} \bibinfo{volume}{5},
  \bibinfo{pages}{19:1--19:19}.
\bibitem[{He and McAuley(2016)}]{DBLP:conf/aaai/HeM16}
\bibinfo{author}{He, R.}, \bibinfo{author}{McAuley, J.J.},
  \bibinfo{year}{2016}.
\newblock \bibinfo{title}{{VBPR:} visual bayesian personalized ranking from
  implicit feedback}, in: \bibinfo{booktitle}{Proceedings of the Thirtieth
  {AAAI} Conference on Artificial Intelligence, February 12-17, 2016, Phoenix,
  Arizona, {USA}}, pp. \bibinfo{pages}{144--150}.
\bibitem[{He et~al.(2018a)He, He, Du and Chua}]{DBLP:conf/sigir/0001HDC18}
\bibinfo{author}{He, X.}, \bibinfo{author}{He, Z.}, \bibinfo{author}{Du, X.},
  \bibinfo{author}{Chua, T.}, \bibinfo{year}{2018}a.
\newblock \bibinfo{title}{Adversarial personalized ranking for recommendation},
  in: \bibinfo{booktitle}{The 41st International {ACM} {SIGIR} Conference on
  Research {\&} Development in Information Retrieval, {SIGIR} 2018, Ann Arbor,
  MI, USA, July 08-12, 2018}, pp. \bibinfo{pages}{355--364}.
\bibitem[{He et~al.(2017)He, Liao, Zhang, Nie, Hu and
  Chua}]{DBLP:conf/www/HeLZNHC17}
\bibinfo{author}{He, X.}, \bibinfo{author}{Liao, L.}, \bibinfo{author}{Zhang,
  H.}, \bibinfo{author}{Nie, L.}, \bibinfo{author}{Hu, X.},
  \bibinfo{author}{Chua, T.}, \bibinfo{year}{2017}.
\newblock \bibinfo{title}{Neural collaborative filtering}, in:
  \bibinfo{booktitle}{Proceedings of the 26th International Conference on World
  Wide Web, {WWW} 2017, Perth, Australia, April 3-7, 2017}, pp.
  \bibinfo{pages}{173--182}.
\bibitem[{He et~al.(2018b)He, Chen, Zhu and Caverlee}]{DBLP:conf/icdm/HeCZC18}
\bibinfo{author}{He, Y.}, \bibinfo{author}{Chen, H.}, \bibinfo{author}{Zhu,
  Z.}, \bibinfo{author}{Caverlee, J.}, \bibinfo{year}{2018}b.
\newblock \bibinfo{title}{Pseudo-implicit feedback for alleviating data
  sparsity in top-k recommendation}, in: \bibinfo{booktitle}{{IEEE}
  International Conference on Data Mining, {ICDM} 2018, Singapore, November
  17-20, 2018}, pp. \bibinfo{pages}{1025--1030}.
\bibitem[{Jang et~al.(2017)Jang, Gu and Poole}]{DBLP:conf/iclr/JangGP17}
\bibinfo{author}{Jang, E.}, \bibinfo{author}{Gu, S.}, \bibinfo{author}{Poole,
  B.}, \bibinfo{year}{2017}.
\newblock \bibinfo{title}{Categorical reparameterization with gumbel-softmax},
  in: \bibinfo{booktitle}{5th International Conference on Learning
  Representations, {ICLR} 2017, Toulon, France, April 24-26, 2017}, pp.
  \bibinfo{pages}{67--77}.
\bibitem[{Kang et~al.(2017)Kang, Fang, Wang and
  McAuley}]{DBLP:conf/icdm/KangFWM17}
\bibinfo{author}{Kang, W.}, \bibinfo{author}{Fang, C.}, \bibinfo{author}{Wang,
  Z.}, \bibinfo{author}{McAuley, J.J.}, \bibinfo{year}{2017}.
\newblock \bibinfo{title}{Visually-aware fashion recommendation and design with
  generative image models}, in: \bibinfo{booktitle}{2017 {IEEE} International
  Conference on Data Mining, {ICDM} 2017, New Orleans, LA, USA, November 18-21,
  2017}, pp. \bibinfo{pages}{207--216}.
\bibitem[{Kingma and Welling(2014)}]{DBLP:journals/corr/KingmaW13}
\bibinfo{author}{Kingma, D.P.}, \bibinfo{author}{Welling, M.},
  \bibinfo{year}{2014}.
\newblock \bibinfo{title}{Auto-encoding variational bayes}, in:
  \bibinfo{booktitle}{2nd International Conference on Learning Representations,
  {ICLR} 2014, Banff, AB, Canada, April 14-16, 2014}, pp.
  \bibinfo{pages}{576--57}.
\bibitem[{Koren et~al.(2009)Koren, Bell and
  Volinsky}]{DBLP:journals/computer/KorenBV09}
\bibinfo{author}{Koren, Y.}, \bibinfo{author}{Bell, R.M.},
  \bibinfo{author}{Volinsky, C.}, \bibinfo{year}{2009}.
\newblock \bibinfo{title}{Matrix factorization techniques for recommender
  systems}.
\newblock \bibinfo{journal}{{IEEE} Computer} \bibinfo{volume}{42},
  \bibinfo{pages}{30--37}.
\bibitem[{Li et~al.(2020)Li, Wu and Wang}]{DBLP:conf/wsdm/LiW020}
\bibinfo{author}{Li, R.}, \bibinfo{author}{Wu, X.}, \bibinfo{author}{Wang, W.},
  \bibinfo{year}{2020}.
\newblock \bibinfo{title}{Adversarial learning to compare: Self-attentive
  prospective customer recommendation in location based social networks}, in:
  \bibinfo{booktitle}{{WSDM} '20: The Thirteenth {ACM} International Conference
  on Web Search and Data Mining, Houston, TX, USA, February 3-7, 2020}, pp.
  \bibinfo{pages}{349--357}.
\bibitem[{Liu et~al.(2019a)Liu, Fu, Qu and Lv}]{DBLP:journals/taslp/LiuFQL19}
\bibinfo{author}{Liu, D.}, \bibinfo{author}{Fu, J.}, \bibinfo{author}{Qu, Q.},
  \bibinfo{author}{Lv, J.}, \bibinfo{year}{2019}a.
\newblock \bibinfo{title}{{BFGAN:} backward and forward generative adversarial
  networks for lexically constrained sentence generation}.
\newblock \bibinfo{journal}{{IEEE/ACM} Trans. Audio, Speech {\&} Language
  Processing} \bibinfo{volume}{27}, \bibinfo{pages}{2350--2361}.
\bibitem[{Liu et~al.(2020)Liu, Pan and Ming}]{DBLP:journals/kbs/LiuPM20}
\bibinfo{author}{Liu, J.}, \bibinfo{author}{Pan, W.}, \bibinfo{author}{Ming,
  Z.}, \bibinfo{year}{2020}.
\newblock \bibinfo{title}{Cofigan: Collaborative filtering by generative and
  discriminative training for one-class recommendation}.
\newblock \bibinfo{journal}{Knowl. Based Syst.} \bibinfo{volume}{191},
  \bibinfo{pages}{105255}.
\bibitem[{Liu et~al.(2019b)Liu, Wang, Yao and Yin}]{DBLP:conf/ijcai/0061W0Y19}
\bibinfo{author}{Liu, W.}, \bibinfo{author}{Wang, Z.}, \bibinfo{author}{Yao,
  B.}, \bibinfo{author}{Yin, J.}, \bibinfo{year}{2019}b.
\newblock \bibinfo{title}{Geo-alm: {POI} recommendation by fusing geographical
  information and adversarial learning mechanism}, in:
  \bibinfo{booktitle}{Proceedings of the Twenty-Eighth International Joint
  Conference on Artificial Intelligence, {IJCAI} 2019, Macao, China, August
  10-16, 2019}, pp. \bibinfo{pages}{1807--1813}.
\bibitem[{Mao et~al.(2017)Mao, Li, Xie, Lau, Wang and
  Smolley}]{DBLP:conf/iccv/MaoLXLWS17}
\bibinfo{author}{Mao, X.}, \bibinfo{author}{Li, Q.}, \bibinfo{author}{Xie, H.},
  \bibinfo{author}{Lau, R.Y.K.}, \bibinfo{author}{Wang, Z.},
  \bibinfo{author}{Smolley, S.P.}, \bibinfo{year}{2017}.
\newblock \bibinfo{title}{Least squares generative adversarial networks}, in:
  \bibinfo{booktitle}{{IEEE} International Conference on Computer Vision,
  {ICCV} 2017, Venice, Italy, October 22-29, 2017}, pp.
  \bibinfo{pages}{2813--2821}.
\bibitem[{Mirza and Osindero(2014)}]{DBLP:journals/corr/MirzaO14}
\bibinfo{author}{Mirza, M.}, \bibinfo{author}{Osindero, S.},
  \bibinfo{year}{2014}.
\newblock \bibinfo{title}{Conditional generative adversarial nets}.
\newblock \bibinfo{journal}{CoRR} \bibinfo{volume}{abs/1411.1784}.
\bibitem[{Nie et~al.(2020)Nie, Wang, Liu, Nie and
  Su}]{DBLP:journals/tomccap/NieWLNS20}
\bibinfo{author}{Nie, W.}, \bibinfo{author}{Wang, W.}, \bibinfo{author}{Liu,
  A.}, \bibinfo{author}{Nie, J.}, \bibinfo{author}{Su, Y.},
  \bibinfo{year}{2020}.
\newblock \bibinfo{title}{{HGAN:} holistic generative adversarial networks for
  two-dimensional image-based three-dimensional object retrieval}.
\newblock \bibinfo{journal}{{TOMM}} \bibinfo{volume}{15},
  \bibinfo{pages}{101:1--101:24}.
\bibitem[{Ouyang et~al.(2014)Ouyang, Liu, Rong and
  Xiong}]{DBLP:conf/iconip/OuyangLRX14}
\bibinfo{author}{Ouyang, Y.}, \bibinfo{author}{Liu, W.}, \bibinfo{author}{Rong,
  W.}, \bibinfo{author}{Xiong, Z.}, \bibinfo{year}{2014}.
\newblock \bibinfo{title}{Autoencoder-based collaborative filtering}, in:
  \bibinfo{booktitle}{Neural Information Processing - 21st International
  Conference, {ICONIP} 2014, Kuching, Malaysia, November 3-6, 2014}, pp.
  \bibinfo{pages}{284--291}.
\bibitem[{Park and Chang(2019)}]{DBLP:conf/www/ParkC19}
\bibinfo{author}{Park, D.H.}, \bibinfo{author}{Chang, Y.},
  \bibinfo{year}{2019}.
\newblock \bibinfo{title}{Adversarial sampling and training for semi-supervised
  information retrieval}, in: \bibinfo{booktitle}{The World Wide Web
  Conference, {WWW} 2019, San Francisco, CA, USA, May 13-17, 2019}, pp.
  \bibinfo{pages}{1443--1453}.
\bibitem[{Perera and Zimmermann(2019)}]{DBLP:conf/www/PereraZ19}
\bibinfo{author}{Perera, D.}, \bibinfo{author}{Zimmermann, R.},
  \bibinfo{year}{2019}.
\newblock \bibinfo{title}{Cngan: Generative adversarial networks for
  cross-network user preference generation for non-overlapped users}, in:
  \bibinfo{booktitle}{The World Wide Web Conference, {WWW} 2019, San Francisco,
  CA, USA, May 13-17, 2019}, pp. \bibinfo{pages}{3144--3150}.
\bibitem[{Quintanilla et~al.(2020)Quintanilla, Rawat, Sakryukin, Shah and
  Kankanhalli}]{DBLP:journals/corr/abs-2004-00698}
\bibinfo{author}{Quintanilla, E.}, \bibinfo{author}{Rawat, Y.S.},
  \bibinfo{author}{Sakryukin, A.}, \bibinfo{author}{Shah, M.},
  \bibinfo{author}{Kankanhalli, M.S.}, \bibinfo{year}{2020}.
\newblock \bibinfo{title}{Adversarial learning for personalized tag
  recommendation}.
\newblock \bibinfo{journal}{CoRR} \bibinfo{volume}{abs/2004.00698}.
\bibitem[{Radford et~al.(2016)Radford, Metz and
  Chintala}]{DBLP:journals/corr/RadfordMC15}
\bibinfo{author}{Radford, A.}, \bibinfo{author}{Metz, L.},
  \bibinfo{author}{Chintala, S.}, \bibinfo{year}{2016}.
\newblock \bibinfo{title}{Unsupervised representation learning with deep
  convolutional generative adversarial networks}, in: \bibinfo{booktitle}{4th
  International Conference on Learning Representations, {ICLR} 2016, San Juan,
  Puerto Rico, May 2-4, 2016}, pp. \bibinfo{pages}{59--66}.
\bibitem[{Rafailidis and Crestani(2019)}]{DBLP:conf/sigir/RafailidisC19}
\bibinfo{author}{Rafailidis, D.}, \bibinfo{author}{Crestani, F.},
  \bibinfo{year}{2019}.
\newblock \bibinfo{title}{Adversarial training for review-based
  recommendations}, in: \bibinfo{booktitle}{Proceedings of the 42nd
  International {ACM} {SIGIR} Conference on Research and Development in
  Information Retrieval, {SIGIR} 2019, Paris, France, July 21-25, 2019}, pp.
  \bibinfo{pages}{1057--1060}.
\bibitem[{Singh and Anand(2019)}]{DBLP:conf/wsdm/SinghA19}
\bibinfo{author}{Singh, J.}, \bibinfo{author}{Anand, A.}, \bibinfo{year}{2019}.
\newblock \bibinfo{title}{{EXS:} explainable search using local model agnostic
  interpretability}, in: \bibinfo{booktitle}{Proceedings of the Twelfth {ACM}
  International Conference on Web Search and Data Mining, {WSDM} 2019,
  Melbourne, VIC, Australia, February 11-15, 2019}, pp.
  \bibinfo{pages}{770--773}.
\bibitem[{Sun et~al.(2020)Sun, Liu, Liu, Ren, Gan and
  Nie}]{DBLP:conf/wsdm/SunLLRGN20}
\bibinfo{author}{Sun, C.}, \bibinfo{author}{Liu, H.}, \bibinfo{author}{Liu,
  M.}, \bibinfo{author}{Ren, Z.}, \bibinfo{author}{Gan, T.},
  \bibinfo{author}{Nie, L.}, \bibinfo{year}{2020}.
\newblock \bibinfo{title}{{LARA:} attribute-to-feature adversarial learning for
  new-item recommendation}, in: \bibinfo{booktitle}{{WSDM} '20: The Thirteenth
  {ACM} International Conference on Web Search and Data Mining, Houston, TX,
  USA, February 3-7, 2020}, pp. \bibinfo{pages}{582--590}.
\bibitem[{Sun et~al.(2019)Sun, Wu, Wu and Ye}]{DBLP:journals/eswa/SunWWY19}
\bibinfo{author}{Sun, Z.}, \bibinfo{author}{Wu, B.}, \bibinfo{author}{Wu, Y.},
  \bibinfo{author}{Ye, Y.}, \bibinfo{year}{2019}.
\newblock \bibinfo{title}{{APL:} adversarial pairwise learning for recommender
  systems}.
\newblock \bibinfo{journal}{Expert Syst. Appl.} \bibinfo{volume}{118},
  \bibinfo{pages}{573--584}.
\bibitem[{Tang et~al.(2012)Tang, Gao and Liu}]{DBLP:conf/wsdm/TangGL12}
\bibinfo{author}{Tang, J.}, \bibinfo{author}{Gao, H.}, \bibinfo{author}{Liu,
  H.}, \bibinfo{year}{2012}.
\newblock \bibinfo{title}{Mtrust: discerning multi-faceted trust in a connected
  world}, in: \bibinfo{booktitle}{Proceedings of the Fifth International
  Conference on Web Search and Web Data Mining, {WSDM} 2012, Seattle, WA, USA,
  February 8-12, 2012}, pp. \bibinfo{pages}{93--102}.
\bibitem[{Tang et~al.(2018)Tang, He, Du, Yuan, Tian and
  Chua}]{DBLP:journals/corr/abs-1809-07062}
\bibinfo{author}{Tang, J.}, \bibinfo{author}{He, X.}, \bibinfo{author}{Du, X.},
  \bibinfo{author}{Yuan, F.}, \bibinfo{author}{Tian, Q.},
  \bibinfo{author}{Chua, T.}, \bibinfo{year}{2018}.
\newblock \bibinfo{title}{Adversarial training towards robust multimedia
  recommender system}.
\newblock \bibinfo{journal}{CoRR} \bibinfo{volume}{abs/1809.07062}.
\bibitem[{Tong et~al.(2019)Tong, Luo, Zhang, Sadiq and
  Cui}]{DBLP:conf/icde/TongLZSC19}
\bibinfo{author}{Tong, Y.}, \bibinfo{author}{Luo, Y.}, \bibinfo{author}{Zhang,
  Z.}, \bibinfo{author}{Sadiq, S.W.}, \bibinfo{author}{Cui, P.},
  \bibinfo{year}{2019}.
\newblock \bibinfo{title}{Collaborative generative adversarial network for
  recommendation systems}, in: \bibinfo{booktitle}{35th {IEEE} International
  Conference on Data Engineering Workshops, {ICDE} Workshops 2019, Macao,
  China, April 8-12, 2019}, pp. \bibinfo{pages}{161--168}.
\bibitem[{Tran et~al.(2019)Tran, Sweeney and Lee}]{DBLP:conf/sigir/TranSL19}
\bibinfo{author}{Tran, T.}, \bibinfo{author}{Sweeney, R.},
  \bibinfo{author}{Lee, K.}, \bibinfo{year}{2019}.
\newblock \bibinfo{title}{Adversarial mahalanobis distance-based attentive song
  recommender for automatic playlist continuation}, in:
  \bibinfo{booktitle}{Proceedings of the 42nd International {ACM} {SIGIR}
  Conference on Research and Development in Information Retrieval, {SIGIR}
  2019, Paris, France, July 21-25, 2019}, pp. \bibinfo{pages}{245--254}.
\bibitem[{Vincent et~al.(2008)Vincent, Larochelle, Bengio and
  Manzagol}]{DBLP:conf/icml/VincentLBM08}
\bibinfo{author}{Vincent, P.}, \bibinfo{author}{Larochelle, H.},
  \bibinfo{author}{Bengio, Y.}, \bibinfo{author}{Manzagol, P.},
  \bibinfo{year}{2008}.
\newblock \bibinfo{title}{Extracting and composing robust features with
  denoising autoencoders}, in: \bibinfo{booktitle}{Machine Learning,
  Proceedings of the Twenty-Fifth International Conference {(ICML} 2008),
  Helsinki, Finland, June 5-9, 2008}, pp. \bibinfo{pages}{1096--1103}.
\bibitem[{Wang et~al.(2019a)Wang, Niepert and
  Li}]{DBLP:journals/corr/abs-1903-10794}
\bibinfo{author}{Wang, C.}, \bibinfo{author}{Niepert, M.}, \bibinfo{author}{Li,
  H.}, \bibinfo{year}{2019}a.
\newblock \bibinfo{title}{Recsys-dan: Discriminative adversarial networks for
  cross-domain recommender systems}.
\newblock \bibinfo{journal}{CoRR} \bibinfo{volume}{abs/1903.10794}.
\bibitem[{Wang et~al.(2019b)Wang, Shao and Lian}]{DBLP:conf/aaai/WangSL19}
\bibinfo{author}{Wang, H.}, \bibinfo{author}{Shao, N.}, \bibinfo{author}{Lian,
  D.}, \bibinfo{year}{2019}b.
\newblock \bibinfo{title}{Adversarial binary collaborative filtering for
  implicit feedback}, in: \bibinfo{booktitle}{The Thirty-Third {AAAI}
  Conference on Artificial Intelligence, {AAAI} 2019, Honolulu, Hawaii, USA,
  January 27 - February 1, 2019}, pp. \bibinfo{pages}{5248--5255}.
\bibitem[{Wang et~al.(2018a)Wang, Wang, Wang, Zhao, Zhang, Zhang, Xie and
  Guo}]{DBLP:conf/aaai/WangWWZZZXG18}
\bibinfo{author}{Wang, H.}, \bibinfo{author}{Wang, J.}, \bibinfo{author}{Wang,
  J.}, \bibinfo{author}{Zhao, M.}, \bibinfo{author}{Zhang, W.},
  \bibinfo{author}{Zhang, F.}, \bibinfo{author}{Xie, X.}, \bibinfo{author}{Guo,
  M.}, \bibinfo{year}{2018}a.
\newblock \bibinfo{title}{Graphgan: Graph representation learning with
  generative adversarial nets}, in: \bibinfo{booktitle}{Proceedings of the
  Thirty-Second {AAAI} Conference on Artificial Intelligence, (AAAI-18), New
  Orleans, Louisiana, USA, February 2-7, 2018}, pp.
  \bibinfo{pages}{2508--2515}.
\bibitem[{Wang et~al.(2017a)Wang, Yu, Zhang, Gong, Xu, Wang, Zhang and
  Zhang}]{DBLP:conf/sigir/WangYZGXWZZ17}
\bibinfo{author}{Wang, J.}, \bibinfo{author}{Yu, L.}, \bibinfo{author}{Zhang,
  W.}, \bibinfo{author}{Gong, Y.}, \bibinfo{author}{Xu, Y.},
  \bibinfo{author}{Wang, B.}, \bibinfo{author}{Zhang, P.},
  \bibinfo{author}{Zhang, D.}, \bibinfo{year}{2017}a.
\newblock \bibinfo{title}{{IRGAN:} {A} minimax game for unifying generative and
  discriminative information retrieval models}, in:
  \bibinfo{booktitle}{Proceedings of the 40th International {ACM} {SIGIR}
  Conference on Research and Development in Information Retrieval, Shinjuku,
  Tokyo, Japan, August 7-11, 2017}, pp. \bibinfo{pages}{515--524}.
\bibitem[{Wang et~al.(2018b)Wang, Yin, Hu, Lian, Wang and
  Huang}]{DBLP:conf/kdd/WangYHLWH18}
\bibinfo{author}{Wang, Q.}, \bibinfo{author}{Yin, H.}, \bibinfo{author}{Hu,
  Z.}, \bibinfo{author}{Lian, D.}, \bibinfo{author}{Wang, H.},
  \bibinfo{author}{Huang, Z.}, \bibinfo{year}{2018}b.
\newblock \bibinfo{title}{Neural memory streaming recommender networks with
  adversarial training}, in: \bibinfo{booktitle}{Proceedings of the 24th {ACM}
  {SIGKDD} International Conference on Knowledge Discovery {\&} Data Mining,
  {KDD} 2018, London, UK, August 19-23, 2018}, pp. \bibinfo{pages}{2467--2475}.
\bibitem[{Wang et~al.(2019c)Wang, Yin, Wang, Nguyen, Huang and
  Cui}]{DBLP:conf/kdd/WangYWNHC19}
\bibinfo{author}{Wang, Q.}, \bibinfo{author}{Yin, H.}, \bibinfo{author}{Wang,
  H.}, \bibinfo{author}{Nguyen, Q.V.H.}, \bibinfo{author}{Huang, Z.},
  \bibinfo{author}{Cui, L.}, \bibinfo{year}{2019}c.
\newblock \bibinfo{title}{Enhancing collaborative filtering with generative
  augmentation}, in: \bibinfo{booktitle}{Proceedings of the 25th {ACM} {SIGKDD}
  International Conference on Knowledge Discovery {\&} Data Mining, {KDD} 2019,
  Anchorage, AK, USA, August 4-8, 2019}, pp. \bibinfo{pages}{548--556}.
\bibitem[{Wang et~al.(2020)Wang, Gong, Wu and
  Zhang}]{DBLP:journals/isci/WangGWZ20}
\bibinfo{author}{Wang, S.}, \bibinfo{author}{Gong, M.}, \bibinfo{author}{Wu,
  Y.}, \bibinfo{author}{Zhang, M.}, \bibinfo{year}{2020}.
\newblock \bibinfo{title}{Multi-objective optimization for location-based and
  preferences-aware recommendation}.
\newblock \bibinfo{journal}{Inf. Sci.} \bibinfo{volume}{513},
  \bibinfo{pages}{614--626}.
\bibitem[{Wang et~al.(2017b)Wang, He, Nie and Chua}]{DBLP:conf/sigir/Wang0NC17}
\bibinfo{author}{Wang, X.}, \bibinfo{author}{He, X.}, \bibinfo{author}{Nie,
  L.}, \bibinfo{author}{Chua, T.}, \bibinfo{year}{2017}b.
\newblock \bibinfo{title}{Item silk road: Recommending items from information
  domains to social users}, in: \bibinfo{booktitle}{Proceedings of the 40th
  International {ACM} {SIGIR} Conference on Research and Development in
  Information Retrieval, Shinjuku, Tokyo, Japan, August 7-11, 2017}, pp.
  \bibinfo{pages}{185--194}.
\bibitem[{Wang et~al.(2019d)Wang, Wang, Xu, He, Cao and
  Chua}]{DBLP:conf/aaai/WangWX00C19}
\bibinfo{author}{Wang, X.}, \bibinfo{author}{Wang, D.}, \bibinfo{author}{Xu,
  C.}, \bibinfo{author}{He, X.}, \bibinfo{author}{Cao, Y.},
  \bibinfo{author}{Chua, T.}, \bibinfo{year}{2019}d.
\newblock \bibinfo{title}{Explainable reasoning over knowledge graphs for
  recommendation}, in: \bibinfo{booktitle}{The Thirty-Third {AAAI} Conference
  on Artificial Intelligence, {AAAI} 2019, Honolulu, Hawaii, USA, January 27 -
  February 1, 2019}, pp. \bibinfo{pages}{5329--5336}.
\bibitem[{Wang et~al.(2018c)Wang, Ou, Tu and Liu}]{DBLP:conf/aims2/WangOTL18}
\bibinfo{author}{Wang, Y.}, \bibinfo{author}{Ou, X.}, \bibinfo{author}{Tu, L.},
  \bibinfo{author}{Liu, L.}, \bibinfo{year}{2018}c.
\newblock \bibinfo{title}{Effective facial obstructions removal with enhanced
  cycle-consistent generative adversarial networks}, in:
  \bibinfo{booktitle}{Artificial Intelligence and Mobile Services - {AIMS} 2018
  - 7th International Conference, Held as Part of the Services Conference
  Federation, {SCF} 2018, Seattle, WA, USA, June 25-30, 2018}, pp.
  \bibinfo{pages}{210--220}.
\bibitem[{Wang et~al.(2019e)Wang, Gao, Wang, Yu, Wen and
  Xiong}]{DBLP:conf/pakdd/WangGWYWX19}
\bibinfo{author}{Wang, Z.}, \bibinfo{author}{Gao, M.}, \bibinfo{author}{Wang,
  X.}, \bibinfo{author}{Yu, J.}, \bibinfo{author}{Wen, J.},
  \bibinfo{author}{Xiong, Q.}, \bibinfo{year}{2019}e.
\newblock \bibinfo{title}{A minimax game for generative and discriminative
  sample models for recommendation}, in: \bibinfo{booktitle}{Advances in
  Knowledge Discovery and Data Mining - 23rd Pacific-Asia Conference, {PAKDD}
  2019, Macau, China, April 14-17, 2019, Proceedings, Part {II}}, pp.
  \bibinfo{pages}{420--431}.
\bibitem[{Williams(1992)}]{DBLP:journals/ml/Williams92}
\bibinfo{author}{Williams, R.J.}, \bibinfo{year}{1992}.
\newblock \bibinfo{title}{Simple statistical gradient-following algorithms for
  connectionist reinforcement learning}.
\newblock \bibinfo{journal}{Machine Learning} \bibinfo{volume}{8},
  \bibinfo{pages}{229--256}.
\bibitem[{Wu et~al.(2017)Wu, Ahmed, Beutel, Smola and
  Jing}]{DBLP:conf/wsdm/WuABSJ17}
\bibinfo{author}{Wu, C.}, \bibinfo{author}{Ahmed, A.}, \bibinfo{author}{Beutel,
  A.}, \bibinfo{author}{Smola, A.J.}, \bibinfo{author}{Jing, H.},
  \bibinfo{year}{2017}.
\newblock \bibinfo{title}{Recurrent recommender networks}, in:
  \bibinfo{booktitle}{Proceedings of the Tenth {ACM} International Conference
  on Web Search and Data Mining, {WSDM} 2017, Cambridge, United Kingdom,
  February 6-10, 2017}, pp. \bibinfo{pages}{495--503}.
\bibitem[{Wu et~al.(2018)Wu, Xia, Tian, Zhao, Qin, Lai and
  Liu}]{DBLP:conf/acml/WuXTZQLL18}
\bibinfo{author}{Wu, L.}, \bibinfo{author}{Xia, Y.}, \bibinfo{author}{Tian,
  F.}, \bibinfo{author}{Zhao, L.}, \bibinfo{author}{Qin, T.},
  \bibinfo{author}{Lai, J.}, \bibinfo{author}{Liu, T.}, \bibinfo{year}{2018}.
\newblock \bibinfo{title}{Adversarial neural machine translation}, in:
  \bibinfo{booktitle}{Proceedings of The 10th Asian Conference on Machine
  Learning, {ACML} 2018, Beijing, China, November 14-16, 2018}, pp.
  \bibinfo{pages}{534--549}.
\bibitem[{Wu et~al.(2019)Wu, Liu, Miao, Zhao, Zhao and
  Guan}]{DBLP:conf/ijcai/WuLMZZG19}
\bibinfo{author}{Wu, Q.}, \bibinfo{author}{Liu, Y.}, \bibinfo{author}{Miao,
  C.}, \bibinfo{author}{Zhao, B.}, \bibinfo{author}{Zhao, Y.},
  \bibinfo{author}{Guan, L.}, \bibinfo{year}{2019}.
\newblock \bibinfo{title}{{PD-GAN:} adversarial learning for personalized
  diversity-promoting recommendation}, in: \bibinfo{booktitle}{Proceedings of
  the Twenty-Eighth International Joint Conference on Artificial Intelligence,
  {IJCAI} 2019, Macao, China, August 10-16, 2019}, pp.
  \bibinfo{pages}{3870--3876}.
\bibitem[{Xiong et~al.(2020)Xiong, Qiao, Han, Xiong, Bu, Li, Yue and
  Yuan}]{DBLP:journals/ijon/XiongQHXBLYY20}
\bibinfo{author}{Xiong, X.}, \bibinfo{author}{Qiao, S.}, \bibinfo{author}{Han,
  N.}, \bibinfo{author}{Xiong, F.}, \bibinfo{author}{Bu, Z.},
  \bibinfo{author}{Li, R.}, \bibinfo{author}{Yue, K.}, \bibinfo{author}{Yuan,
  G.}, \bibinfo{year}{2020}.
\newblock \bibinfo{title}{Where to go: An effective point-of-interest
  recommendation framework for heterogeneous social networks}.
\newblock \bibinfo{journal}{Neurocomputing} \bibinfo{volume}{373},
  \bibinfo{pages}{56--69}.
\bibitem[{Yang et~al.(2018)Yang, Guo, Wang, Jiang, Xiao and
  Wang}]{DBLP:conf/icdm/YangGWJXW18}
\bibinfo{author}{Yang, D.}, \bibinfo{author}{Guo, Z.}, \bibinfo{author}{Wang,
  Z.}, \bibinfo{author}{Jiang, J.}, \bibinfo{author}{Xiao, Y.},
  \bibinfo{author}{Wang, W.}, \bibinfo{year}{2018}.
\newblock \bibinfo{title}{A knowledge-enhanced deep recommendation framework
  incorporating gan-based models}, in: \bibinfo{booktitle}{{IEEE} International
  Conference on Data Mining, {ICDM} 2018, Singapore, November 17-20, 2018}, pp.
  \bibinfo{pages}{1368--1373}.
\bibitem[{Yoo et~al.(2017)Yoo, Ha, Yi, Ryu, Kim, Ha, Kim and
  Yoon}]{DBLP:journals/corr/YooHYRKH0Y17}
\bibinfo{author}{Yoo, J.}, \bibinfo{author}{Ha, H.}, \bibinfo{author}{Yi, J.},
  \bibinfo{author}{Ryu, J.}, \bibinfo{author}{Kim, C.}, \bibinfo{author}{Ha,
  J.}, \bibinfo{author}{Kim, Y.}, \bibinfo{author}{Yoon, S.},
  \bibinfo{year}{2017}.
\newblock \bibinfo{title}{Energy-based sequence gans for recommendation and
  their connection to imitation learning}.
\newblock \bibinfo{journal}{CoRR} \bibinfo{volume}{abs/1706.09200}.
\bibitem[{Yu et~al.(2018)Yu, Gao, Li, Yin and Liu}]{DBLP:conf/cikm/Yu0LYL18}
\bibinfo{author}{Yu, J.}, \bibinfo{author}{Gao, M.}, \bibinfo{author}{Li, J.},
  \bibinfo{author}{Yin, H.}, \bibinfo{author}{Liu, H.}, \bibinfo{year}{2018}.
\newblock \bibinfo{title}{Adaptive implicit friends identification over
  heterogeneous network for social recommendation}, in:
  \bibinfo{booktitle}{Proceedings of the 27th {ACM} International Conference on
  Information and Knowledge Management, {CIKM} 2018, Torino, Italy, October
  22-26, 2018}, pp. \bibinfo{pages}{357--366}.
\bibitem[{Yu et~al.(2019a)Yu, Gao, Yin, Li, Gao and
  Wang}]{DBLP:journals/corr/abs-1909-03529}
\bibinfo{author}{Yu, J.}, \bibinfo{author}{Gao, M.}, \bibinfo{author}{Yin, H.},
  \bibinfo{author}{Li, J.}, \bibinfo{author}{Gao, C.}, \bibinfo{author}{Wang,
  Q.}, \bibinfo{year}{2019}a.
\newblock \bibinfo{title}{Generating reliable friends via adversarial training
  to improve social recommendation}.
\newblock \bibinfo{journal}{CoRR} \bibinfo{volume}{abs/1909.03529}.
\bibitem[{Yu et~al.(2019b)Yu, Zhang, Cao and Xia}]{DBLP:conf/ijcai/YuZCX19}
\bibinfo{author}{Yu, X.}, \bibinfo{author}{Zhang, X.}, \bibinfo{author}{Cao,
  Y.}, \bibinfo{author}{Xia, M.}, \bibinfo{year}{2019}b.
\newblock \bibinfo{title}{{VAEGAN:} {A} collaborative filtering framework based
  on adversarial variational autoencoders}, in: \bibinfo{booktitle}{Proceedings
  of the Twenty-Eighth International Joint Conference on Artificial
  Intelligence, {IJCAI} 2019, Macao, China, August 10-16, 2019}, pp.
  \bibinfo{pages}{4206--4212}.
\bibitem[{Yuan et~al.(2019a)Yuan, Yao and
  Benatallah}]{DBLP:conf/ijcnn/YuanYB19}
\bibinfo{author}{Yuan, F.}, \bibinfo{author}{Yao, L.},
  \bibinfo{author}{Benatallah, B.}, \bibinfo{year}{2019}a.
\newblock \bibinfo{title}{Adversarial collaborative auto-encoder for top-n
  recommendation}, in: \bibinfo{booktitle}{International Joint Conference on
  Neural Networks, {IJCNN} 2019 Budapest, Hungary, July 14-19, 2019}, pp.
  \bibinfo{pages}{1--8}.
\bibitem[{Yuan et~al.(2019b)Yuan, Yao and
  Benatallah}]{DBLP:conf/sigir/YuanYB19}
\bibinfo{author}{Yuan, F.}, \bibinfo{author}{Yao, L.},
  \bibinfo{author}{Benatallah, B.}, \bibinfo{year}{2019}b.
\newblock \bibinfo{title}{Adversarial collaborative neural network for robust
  recommendation}, in: \bibinfo{booktitle}{Proceedings of the 42nd
  International {ACM} {SIGIR} Conference on Research and Development in
  Information Retrieval, {SIGIR} 2019, Paris, France, July 21-25, 2019}, pp.
  \bibinfo{pages}{1065--1068}.
\bibitem[{Zhang et~al.(2018)Zhang, Zhang, Zhang and
  Wang}]{DBLP:journals/kbs/ZhangZZW18}
\bibinfo{author}{Zhang, F.}, \bibinfo{author}{Zhang, Z.},
  \bibinfo{author}{Zhang, P.}, \bibinfo{author}{Wang, S.},
  \bibinfo{year}{2018}.
\newblock \bibinfo{title}{{UD-HMM:} an unsupervised method for shilling attack
  detection based on hidden markov model and hierarchical clustering}.
\newblock \bibinfo{journal}{Knowl.-Based Syst.} \bibinfo{volume}{148},
  \bibinfo{pages}{146--166}.
\bibitem[{Zhang(2018)}]{DBLP:conf/sigir/Zhang18}
\bibinfo{author}{Zhang, W.}, \bibinfo{year}{2018}.
\newblock \bibinfo{title}{Generative adversarial nets for information
  retrieval: Fundamentals and advances}, in: \bibinfo{booktitle}{The 41st
  International {ACM} {SIGIR} Conference on Research {\&} Development in
  Information Retrieval, {SIGIR} 2018, Ann Arbor, MI, USA, July 08-12, 2018},
  pp. \bibinfo{pages}{1375--1378}.
\bibitem[{Zhao et~al.(2014)Zhao, McAuley and King}]{DBLP:conf/cikm/ZhaoMK14}
\bibinfo{author}{Zhao, T.}, \bibinfo{author}{McAuley, J.J.},
  \bibinfo{author}{King, I.}, \bibinfo{year}{2014}.
\newblock \bibinfo{title}{Leveraging social connections to improve personalized
  ranking for collaborative filtering}, in: \bibinfo{booktitle}{Proceedings of
  the 23rd {ACM} International Conference on Conference on Information and
  Knowledge Management, {CIKM} 2014, Shanghai, China, November 3-7, 2014}, pp.
  \bibinfo{pages}{261--270}.
\bibitem[{Zhao et~al.(2018)Zhao, Wang, Ye, Gao, Yang and
  Chen}]{DBLP:conf/ijcai/ZhaoWYGYC18}
\bibinfo{author}{Zhao, W.}, \bibinfo{author}{Wang, B.}, \bibinfo{author}{Ye,
  J.}, \bibinfo{author}{Gao, Y.}, \bibinfo{author}{Yang, M.},
  \bibinfo{author}{Chen, X.}, \bibinfo{year}{2018}.
\newblock \bibinfo{title}{{PLASTIC:} prioritize long and short-term information
  in top-n recommendation using adversarial training}, in:
  \bibinfo{booktitle}{Proceedings of the Twenty-Seventh International Joint
  Conference on Artificial Intelligence, {IJCAI} 2018, Stockholm, Sweden, July
  13-19, 2018}, pp. \bibinfo{pages}{3676--3682}.
\bibitem[{Zheng et~al.(2019)Zheng, Song, Chen, Hu, Cao and
  Nie}]{DBLP:conf/mm/ZhengSCHCN19}
\bibinfo{author}{Zheng, N.}, \bibinfo{author}{Song, X.}, \bibinfo{author}{Chen,
  Z.}, \bibinfo{author}{Hu, L.}, \bibinfo{author}{Cao, D.},
  \bibinfo{author}{Nie, L.}, \bibinfo{year}{2019}.
\newblock \bibinfo{title}{Virtually trying on new clothing with arbitrary
  poses}, in: \bibinfo{booktitle}{Proceedings of the 27th {ACM} International
  Conference on Multimedia, {MM} 2019, Nice, France, October 21-25, 2019}, pp.
  \bibinfo{pages}{266--274}.
\bibitem[{Zhou et~al.(2019)Zhou, Yin, Zhang, Trajcevski, Zhong and
  Wu}]{DBLP:conf/www/0002YZTZW19}
\bibinfo{author}{Zhou, F.}, \bibinfo{author}{Yin, R.}, \bibinfo{author}{Zhang,
  K.}, \bibinfo{author}{Trajcevski, G.}, \bibinfo{author}{Zhong, T.},
  \bibinfo{author}{Wu, J.}, \bibinfo{year}{2019}.
\newblock \bibinfo{title}{Adversarial point-of-interest recommendation}, in:
  \bibinfo{booktitle}{The World Wide Web Conference, {WWW} 2019, San Francisco,
  CA, USA, May 13-17, 2019}, pp. \bibinfo{pages}{3462--34618}.

\end{thebibliography}

\printcredits
\end{document}